\documentclass[11pt]{article} 
\usepackage[textheight=216mm,textwidth=160mm]{geometry}
\usepackage{graphicx} 
\usepackage{natbib}
\usepackage{enumerate}

\usepackage{multirow}
\usepackage{makecell}
\usepackage{hyperref}
\hypersetup{
    colorlinks=true,
    linkcolor=blue,
    filecolor=blue,      
    urlcolor=blue,
    citecolor=blue
    }

\usepackage{newtxtext}
\usepackage{newtxmath}

\title{Linking emitted drops to collective bursting bubbles across a wide range of bubble size distributions}

\author{Megan Mazzatenta$^1$, Martin A. Erinin$^1$, Baptiste N\'eel$^1$, Luc Deike$^{1,2}$\thanks{email for correspondance: ldeike@princeton.edu}}

\date{$^1$ Department of Mechanical and Aerospace Engineering, Princeton University, Princeton, NJ, 08544, USA\\
$^2$High Meadows Environmental Institute, Princeton University, Princeton, NJ, 08544, USA}

\begin{document}
\maketitle

\begin{abstract}

Bubbles entrained by breaking waves rise to the ocean surface, where they cluster
before bursting and release droplets into the atmosphere. The ejected drops and dry aerosol particles, left behind after the liquid drop evaporates, affect the radiative balance of the atmosphere and can act as cloud condensation nuclei. The remaining uncertainties surrounding the sea spray emissions function motivate controlled laboratory experiments that directly measure and link collective bursting bubbles and the associated drops and sea salt aerosols. We perform experiments in artificial seawater for a wide range of bubble size distributions, measuring both bulk and surface bubble distributions (measured radii from 30~\textmu{m} to 5~mm), together with the associated drop size distribution (salt aerosols and drops of measured radii from 50~nm to 500~\textmu{m}) to quantify the link between emitted drops and bursting surface bubbles. We evaluate how well the individual bubble bursting scaling laws describe our data across all scales and demonstrate that the measured drop production by collective bubble bursting can be represented by a single framework integrating individual bursting scaling laws over the various bubble sizes present in our experiments. We show that film drop production by bubbles between 100~\textmu{m} and 1~mm describes the submicron drop production, while jet drop production by bubbles from 30~\textmu{m} to 2~mm describes the production of drops larger than 1~\textmu{m}. Our work confirms that sea spray emissions functions based on individual bursting processes are reasonably accurate as long as the surface bursting bubble size distribution is known.

\end{abstract}

\section{Introduction}
When a wave breaks at sea, it entrains air bubbles below the ocean surface. Those bubbles rise back to the surface and reside for some time before bursting and ejecting drops into the atmosphere. The ejected drops can then evaporate, leaving behind ocean spray aerosol particles composed of salts, biological material, and other chemicals found in the ocean. In the atmosphere, these emitted drops and corresponding dry particles affect the radiative balance and can serve as cloud condensation nuclei \citep{lewis_sea_2004,de_leeuw_production_2011,veron_ocean_2015,cochran_sea_2017}. 

One main difficulty of studying spray generation by bubble bursting is the range of scales involved in the related processes, from the single breaking wave of \textit{O}(1\textendash10~m) to entrained bubbles of sizes \textit{O}(0.01\textendash10~mm) and emitted drops of sizes \textit{O}(0.01\textendash100~\textmu{m}). To predict ocean spray generation from knowledge of the bubble size distribution, we must understand the physical link between underwater bubbles that rise to the ocean surface, surface bubbles that cluster and burst, and the resulting drops they eject (depicted in figure 1 of \citet{deike_rev_2022}).  

Parameterizations of the sea spray generation function used in large-scale models (as reviewed by \citet{lewis_sea_2004,de_leeuw_production_2011,veron_ocean_2015}), display large uncertainties of at least an order of magnitude across the full range of emitted drop sizes. These uncertainties come from the difficulty of characterizing the drop formation across many processes and scales, as well as from the various empirical ways the parameterizations are developed from interpretation of field measurements or upscaling of laboratory data. In practice, the uncertainties lead to very different sensitivity to physical parameters such as temperature \citep{forestieri_temperature_2018} and directly impact the modeled radiative balance and climate sensitivity in climate models \citep{paulot_revisiting_2020}.

Individual bubble bursting has been extensively studied to characterize the production mechanism, size, and number of drops produced by the bursting of a single bubble. Previous laboratory experiments have focused either on film drops \citep{blanchard_film_1988,resch_film_1991,resch_submicron_1992,spiel_film_97,spiel_film_98,lv2012,jiang2022} or jet drops \citep{spiel_film_97,spiel_film_98,ghabache_physics_2014,ghabache_size_2016,walls_jet_2015,brasz_minimum_2018}, and numerical simulations have studied jet drop production \citep{deike_dynamics_2018,brasz_minimum_2018,berny_role_2020,berny2021}. From those studies, scaling laws have been developed that relate the bursting bubble size to the number and size of drops emitted through jet drop production \citep{ganan-calvo_revision_2017, ganan-calvo_ocean_2023, blancorodriguez_sea_2020}, film drop production through a centrifuge mechanism \citep{lv2012}, and film drop production through a proposed flapping mechanism \citep{jiang2022,jiang_abyss_2024}. However, the extent to which these scaling laws, developed for idealized individual bubble bursting, apply to more realistic configurations like clusters of bubbles within ocean whitecaps, remains to be experimentally demonstrated. 

Combining the above-mentioned individual bubble bursting scaling laws for the various mechanisms with the bubble size distribution under a breaking wave and the statistics of air entrainment at large scale, \citet{deike_mechanistic_2022} developed a mechanistic sea spray emissions function and demonstrated the remarkable coherence with empirically-derived emissions functions. To obtain the ocean spray emissions function, \citet{deike_mechanistic_2022} integrates over a given bubble size distribution (constrained by laboratory and numerical studies by \citet{deane_scale_2002,deike_air_2016,gao_shen_2021,mostert_high-resolution_2022}), assuming individual bubble bursting scaling laws for the size, number, and distribution of emitted drops for each production mechanism. The method uses the integration introduced for film drop production by \citet{lv2012} and extended to jet drop production in \citet{berny2021}, and air entrainment volume by breaking waves statistics in the ocean \citep{deike_air_2016,deike_air_2017}. 

The approach from \citet{deike_mechanistic_2022}, as well as past approaches (see the overview in \citet{lewis_sea_2004}), assumes that the bubble size distribution under a breaking wave is always of the same shape and does not consider collective bubble bursting or surfactant effects. Physico-chemical properties of the ocean water \textendash\ such as salinity, temperature, surfactant concentration, and biological composition \textendash\ bring another layer of complexity to sea spray generation by potentially modulating the bubble processes (i.e. lifetime, coalescence, bursting). 
Modeling of sea spray organic aerosol emissions in \citet{burrows_oceanfilms_2014} highlights the numerous chemical and biological factors that are thought to influence spray generation, showing that \textendash\ despite extensive past experimental efforts \textendash\ there is still a need for controlled, well-constrained experiments to improve our knowledge of how these realistic factors control the bursting processes while limiting experimental uncertainties.

To investigate the collective bursting relevant to the open ocean context, previous experiments have measured drop production by collections of bubbles generated through different methods in a variety of complex solutions, focusing especially on the measurement of submicron drops/aerosols. The experiments include bursting bubbles generated by both a plunging sheet and nucleation bubbler in natural and artificial seawater \citep{wang2017}, bursting bubble rafts produced in sampled natural seawater through a forced-air venturi channel \citep{frossard_marine_2019}, the investigation of the effect of phytoplankton and bacteria on aerosol production for bubbles generated by laboratory breaking waves in natural seawater \citep{prather2013}, as well as several other experiments involving bursting bubble rafts or foams including those in solutions of different organic content, or different salt concentration \citep{cipriano_bubble_1981,martensson_laboratory_2003,l-b1_2003,sellegri_surfactants_2006,tyree_foam_2007,modini_effect_2013,l-b2_2017,zinke_effect_2022,dubitsky_effects_2023}. These studies suggest that ocean physico-chemical parameters modulate the bursting process. However, the complexity of the seawater solution and the lack of complete bulk bubble, surface bubble, and drop/dry particle measurements across a wide range of scales make it difficult to compare results from separate studies and to distinguish the specific effects of individual physico-chemical variables among other differences due to varied bubble size distributions or collective bursting. 

These uncertainties and their practical consequences on ocean spray emissions functions motivate an intermediate-scale collective bursting laboratory experiment, measuring bubbles in the bulk and at the surface as well as the produced drops, as a way to bridge the idealized and the oceanic spray generation configurations. Using a well-characterized setup with measurements that capture the full size range of bubbles and associated ejected drops, we can maintain precise control over the solution while observing the effect of any changes in the properties on the collective bursting. This idea was introduced through the experiments of \citet{neel2021,neel_surf_2022}, where systematic measurements of millimetric bubbles and supermicron drops were made for solutions of different surfactant concentrations. 

\begin{figure}[!h]
  \centerline{\includegraphics[scale=0.78]{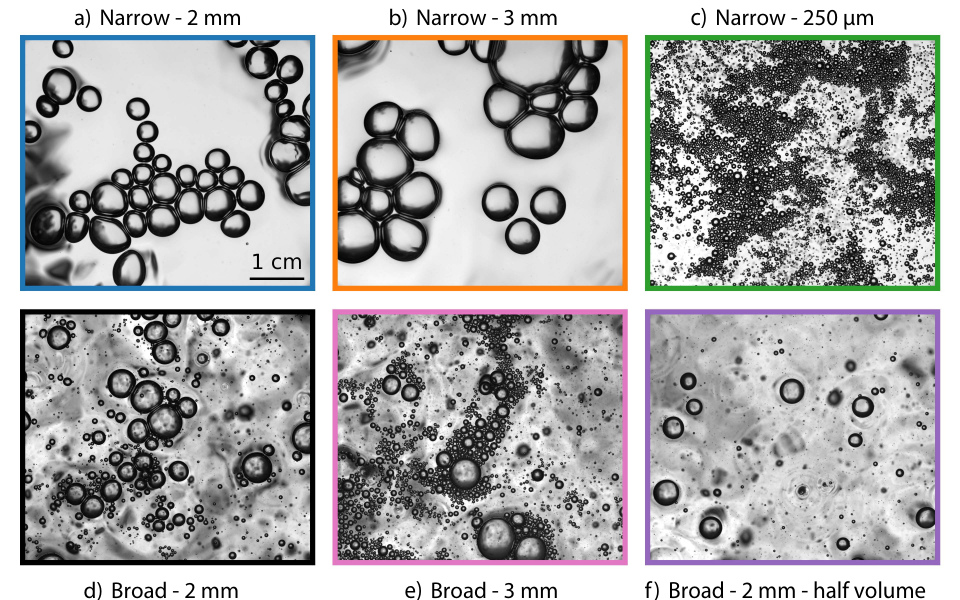}}
  \caption{Characteristic images of bubbles on the water surface for six conditions presented in this paper. The top row shows narrower size distributions of clustered bubbles for injection sizes (a) 2~mm, (b) 3~mm, and (c) 250~\textmu{m}. The bottom row shows broad-banded distributions initially injected as (d, f) 2~mm and (e) 3~mm bubbles before underwater turbulence caused the breakup of these injected bubbles into those spanning a wide range of scales. In (f) half the volume of 2~mm bubbles were injected compared to the other cases (using 16 instead of 32 needles), leading to bubbles that are more isolated instead of clustered at the surface. Details of how bubbles were generated for each case are found in table \ref{tab:cases}.} 
\label{fig:surfims}
\end{figure}

In this paper, we perform experiments with measurements of all relevant scales of bulk and surface bubbles (radii 30~\textmu{m} to 5~mm), liquid drops (radii 10 to 500~\textmu{m}), and salt aerosols (equivalent drop radii 50~nm to 10~\textmu{m}) for various clustered bubble configurations, allowing us to directly characterize the link between the drop size distribution and the surface bursting bubble distribution. We have designed a bubbling tank experiment (building off the work from \citet{neel2021}) where bubbles are generated at the bottom of the tank, and statistical measurements are made in the bulk and at the surface. Three measurement techniques are combined to measure drop sizes spanning four orders of magnitude: optical digital inline holography allows the measurement of drops ejected in the air with radii from 10~\textmu{m} to 500~\textmu{m}, while smaller droplets (not accessible with direct imaging techniques) are inferred by drying and measuring the size of the resulting salt crystals, spanning equivalent drop radii from 50~nm to 10~\textmu{m}. 
We characterize the spray generation by clusters of bubbles in a precisely-controlled solution of artificial seawater, and we vary the bulk bubble size distribution across cases. The bulk bubble size distributions tested include quasi-monodisperse plumes of rising bubbles centered around 3~mm, 2~mm, and 250~\textmu{m}, as well as broad-banded distributions where bubbles range from 30~\textmu{m} to 3~mm. The broad-banded size distributions mimic the bubble size distribution entrained by breaking waves. As illustrated in figure 1, these bubble size distributions exhibit very different characteristics at the surface, which will control the resulting drop size distribution.
We then explore whether we can describe the associated spray generation by these various bubble distributions within a single physical framework.

The paper is organized as follows. We detail the setup, experimental methods, and resulting size distribution data in section \ref{sec:start}. In section \ref{sec:bulksurf}, we discuss linking the bulk bubble and surface bubble measurements, followed by the section \ref{sec:surfdrop} analysis linking the surface bubble and drop measurements. Section \ref{sec:concl} presents the conclusions.

\section{Measuring bubbles, drops, and dry particles in a bubbling tank}
\label{sec:start}
We present the experimental methods, including a description of the bubbling tank, preparation of the artificial seawater solution, measurements of bulk bubbles, surface bubbles, and finally drops/dry particles. Sample size distributions from these measurements are then shown and discussed.

\subsection {Experimental setup and conditions}
\label{setup}

\begin{figure}
  \centerline{\includegraphics[scale=1.3]{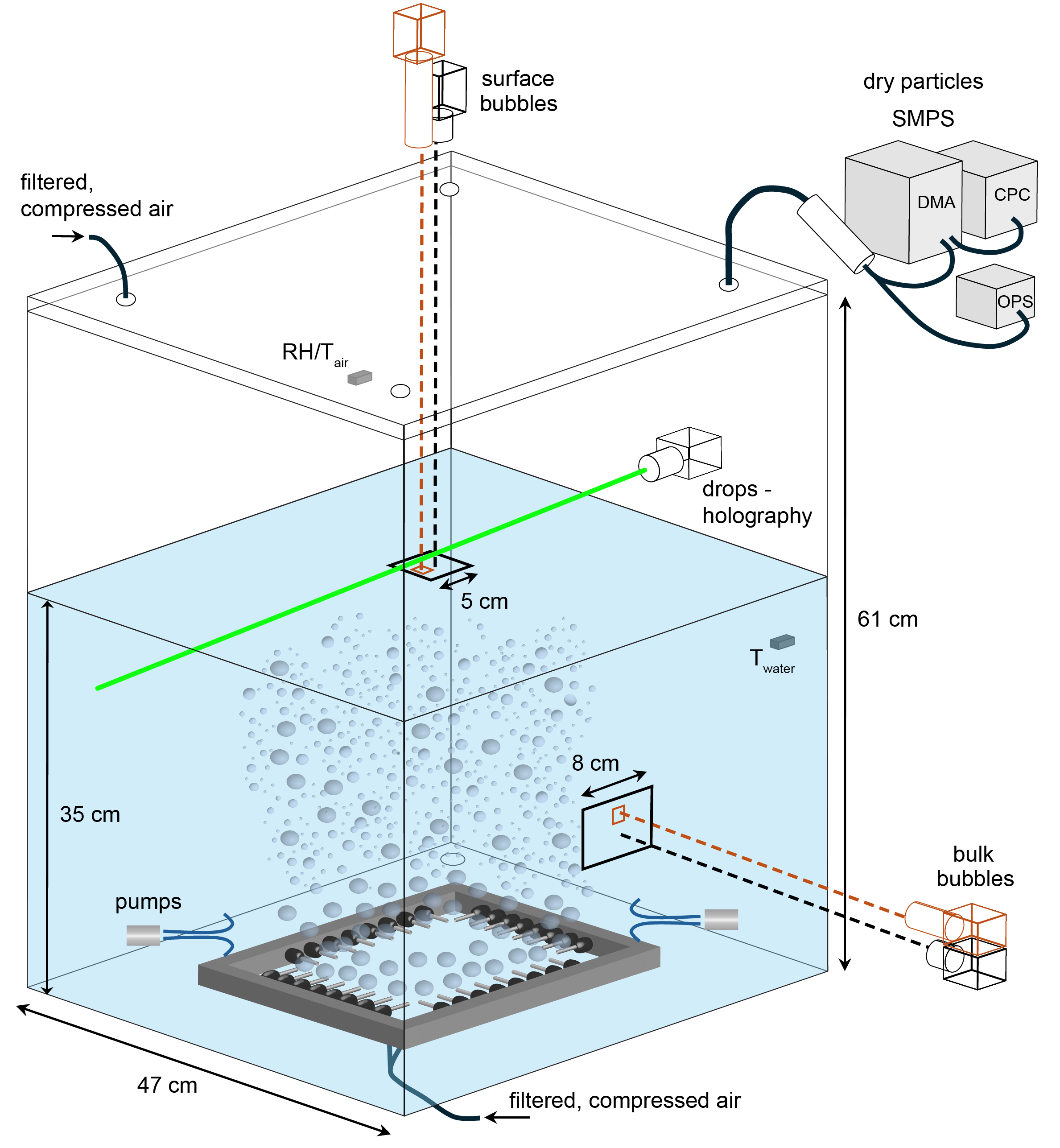}}
  \caption{Sketch of the experimental setup, with measurements of bubbles, drops, and dry aerosol particles. Bubbles of nearly identical sizes are generated by compressed air flowing through needles at the bottom of the tank. When two underwater pumps are turned on (as illustrated in the sketch), the injected bubbles are broken up into many smaller bubbles as they rise into the region of turbulence generated by the water jets from the pumps (bubbles not drawn to scale). The bubbles are imaged from the side view to measure the rising bulk bubbles, and from the top view to capture the bubbles at the surface. Large fields of view (black) and small fields of view (orange) are drawn approximately to scale and permit measurement of bubble radii from 30~\textmu{m} to 5~mm. Liquid drops ejected into the air by bursting bubbles are measured through an inline holographic setup (measuring liquid drop radii from 10\textmu{m} to 500~\textmu{m}), while dry aerosol particles are measured by extracting and drying the liquid drops and analyzing them using an Optical Particle Sizer and Scanning Mobility Particle Sizer (OPS, SMPS composed of DMA and CPC - TSI, Inc.) (for dry particle radii from 20~nm to 5~\textmu{m}). A line of filtered, compressed air is connected to the lid to flush the air above the water surface before the start of each run. Relative humidity and both air and water temperature are monitored throughout the experiments.} 
\label{fig:setup}
\end{figure}

The experimental setup shown in figure \ref{fig:setup}, consists of a polycarbonate tank with inner dimensions 47 x 47 x 61 cm$^3$, similar to the one used by \citet{neel2021}, which is filled to a height of 35~cm with a solution of artificial seawater (made of deionized (DI) water and artificial sea salt, ASTM D1141-98, Lake Products Company LLC). The water height is chosen so that the waves/fluctations on the water surface remain small and the surface bubbles can be clearly imaged. Bubbles are continuously generated as compressed air flows through the needles positioned at the bottom of the tank, arranged in a square with eight needles extending horizontally from each side. The air flowrate through the needles is controlled at 100~cm$^3$/min per needle, so that the resulting plume of rising bubbles is fairly monodisperse. 

The artificial seawater solution is prepared by mixing DI water with ASTM D1141-98 Artificial Sea Salt, which contains the same proportions of elements found in natural seawater in amounts greater than 0.0004\% by weight (Lake Products Company LLC). Solutions discussed in this paper were prepared with a salinity of 36~g/L (35~g/kg) to mimic the composition of ocean water. The solution has liquid viscosity $\mu$ = 1.001 mPa s, liquid density $\rho$ = 1025~kg/m$^3$, and static surface tension $\gamma$ = 52~mN/m. The low static surface tension indicates that the salt mixture likely includes some surfactant. At the beginning and end of each day of experiments, the solution was characterized by measuring the surface tension isotherm with a Langmuir Trough (KSV NIMA, model KN 1003). This measurement was used to confirm that the surface properties of the solution remained consistent over multiple days of experiments. 
Two pumps (Rule 20DA 800 GPH Bilge Pump) are mounted 6~cm above the needle tips, pointed towards each other from opposite corners. When the pumps are turned on, they produce water jets that meet in the center of the tank, creating a region of turbulent flow that causes the rising bubbles to fragment into many smaller bubbles, with turbulence similar to the setup described in \citet{ruth_experimental_2022}. Experiments can be performed with or without forced turbulence, where the turbulence has two effects: agitate the flow and break the bubbles, leading to a broad-banded size distribution. Note that bubble imaging occurs above the region of active breakup.

Three experimental parameters were varied to create different initial bubble size distributions for the different cases run in this experiment: turning the underwater pumps off or on to create either a narrow- or broad-banded bulk bubble distribution, changing the needle size/bubble production method, and reducing the number of needles (which effectively reduces the injected air volume per unit time because the flowrate per needle is kept constant). An aquarium bubbler is also used to generate a plume of submillimetric bubbles centered around 250~\textmu{m}. 

\begin{table}
  \begin{center}
\def~{\hphantom{0}}
  \begin{tabular}{l@{}c@{}c@{}ccccccc}
      Case Description  &  Bubble & Total Air & Needle ID & Pumps & \# Needles & $R_{b,inj}$ & $N_{R_b<2mm}$  \\[0.25pt]
        &  Production & Flowrate & (\textmu{m}) &  & & (mm) & (cm$^{-3}$) \\ [0.25pt]
        &  Method & (cm$^3$/min) &  &  & & &  \\ [6pt]
       Narrow - 2~mm & Needles & 3200 & 432 & off & 32 & 2 & 0.40\\
       Broad - 2~mm & Needles & 3200 &  432 & on & 32 & 2 & 45 \\
       Broad - 2~mm - half volume & Needles & 1600 &  432 & on & 16 & 2 & 25\\
       Narrow - 3~mm & Needles & 3200 &  965 & off & 16 & 3 & 0.13\\
       Broad - 3~mm & Needles & 3200 & 965 & on & 32 & 3 & 43\\
       Narrow - 250~\textmu{m} & Porous Media & 800 & & off &  & 0.25 & 57 \\[1pt] 
        & (Bubbler)  & &  &  & & &  \\

  \end{tabular}
  \caption{Experimental parameters. The bulk bubble size distribution across cases was varied by changing between quiescent and turbulent flow underwater (by turning the underwater pumps off or on, causing the bubbles to break up in the turbulent case), by changing the injection bubble size (using two different needle sizes or an aquarium bubbler), and by halving the volume (using 16 instead of 32 needles with the same air flowrate per needle). Case names describe the range of radii over which a significant number of bubbles are concentrated (narrow, broad) and the initial injection bubble size before any breakup in turbulence (2~mm, 3~mm, 250~\textmu{m}). The Broad-2~mm-half volume case is characterized by bubbles arriving at the surface isolated from each other, bursting individually instead of clustered as the smaller number of needles leads to both reduced volume and increased needle spacing. $R_{b, inj}$ represents the typical injection bubble size by the needles or bubbler. $N_{R_b<2mm}$ shows the number of bubbles (per unit volume) smaller  than radius 2~mm, which reinforces the designation of certain cases as narrow- or broad-banded. We provide the total air flowrate through the needles for each case, but we caution that this flowrate value should not be used to approximate the bubble population in our configuration, as it is primarily related to the volume contained in the largest bubbles and does not capture the variations in the bubble size distributions (broad and narrow configurations) necessary when linking bursting bubbles to emitted drops. Bubble size distributions for each case are shown throughout sections \ref{sec:dist}, \ref{sec:direct}, and \ref{section:forward}. Representative surface bubble images for each case are shown in figure \ref{fig:surfims}.} 
  \label{tab:cases}
  \end{center}
\end{table}

Figure \ref{fig:surfims} illustrates the wide range of surface bubble populations explored in this paper, showing representative images of the various conditions detailed in table \ref{tab:cases}. The surface bubble distributions display marked differences in terms of both sizes and clustering, which will control the emitted drop size distribution. There are also significant differences in the number of bubbles present at the surface in each condition, which are associated with very different numbers of produced drops (see appendix \ref{appA} for the total number of drops compared to number of bubbles for each case). Because of the variety in size, number, and clustering of surface bubbles for various conditions, it is necessary to characterize the surface properly through many measurements, across all scales, before attempting to link the bursting surface bubbles to the drops they produce.  

In order to capture the large range of length scales present in the system \textendash\ spanning bubble radii of 30 to 5000~\textmu{m} and drop sizes of 0.05 to 500~\textmu{m} \textendash\ we make multiple measurements of bulk bubbles, surface bubbles, drops, and dry particles. Bulk bubbles are imaged in both a large and small field of view by cameras positioned above the region of active breakup. Surface bubbles are also captured in two fields-of-view by cameras positioned perpendicular to the water surface, where the bubbles cluster near the center of the tank  before bursting and ejecting drops into the air above the water surface, which is enclosed in the tank with an acrylic lid. 

Drops are sized using a combination of an inline holographic system positioned just above the water surface \citep{erinin, methods}, an Optical Particle Sizer (OPS: TSI, Inc. OPS 3330), and a Scanning Mobility Particle Sizer  (SMPS, consists of: TSI, Inc. Advanced Aerosol Neutralizer, Model 3088; Electrostatic Classifier, Model 3082; Differential Mobility Analyzer (DMA), Model 3081; Condensation Particle Counter (CPC), Model 3752). The SMPS operates with an aerosol flowrate of 0.3~L/min and sheath flowrate of 2.4~L/min. The OPS and SMPS are connected through the lid and pull air with suspended drops out of the tank, where they are dried (TSI Diffusion Dryer, Model 3062) and then measured. Relative humidity and air temperature are monitored 25~cm above the calm water surface throughout all experimental runs, along with the temperature of the solution.

\begin{figure}
  \centerline{\includegraphics[scale=0.6]{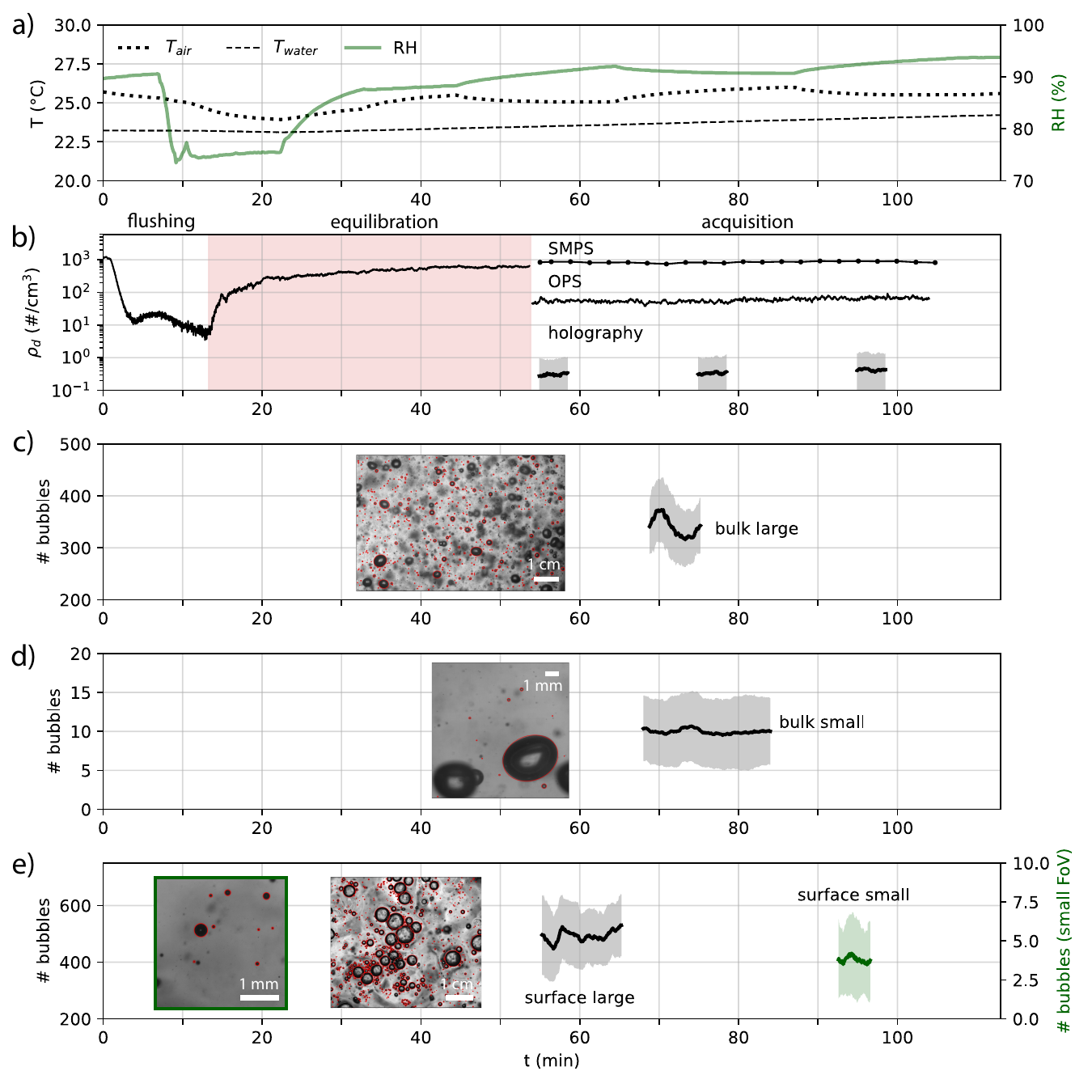}}
  \caption{Experimental protocol for data acquisition illustrated by typical time series of temperature, relative humidity, bubbles, drops, and dry particles through a single run on a representative case (Broad-2~mm in table \ref{tab:cases}). Sample images are shown with bubble detection overlaid in red (c-e). Scale bars (shown as white lines on images) represent 1~cm for the bulk and surface large fields of view (c, e-right image), and 1~mm for the small fields of view (d, e-left image). Panel (a) illustrates the steps in the experimental protocol, showing first flushing to obtain clean air, followed by equilibration during which the relative humidity reaches a value of about 90\%, followed by a steady state. Air and water temperature are also shown in (a). The dry aerosol particle counts are shown in (b), with the SMPS particle count close to 0 (less than 10 particles per cm$^3$) at the end of the flushing stage, and reaching a steady state during the acquisition stage. Drop measurements from the holography are shown in (b) by a concentration $\rho_d$ (number per measurement volume), with the rolling mean and standard deviation plotted for each holographic measurement. For the bubble measurements in panels (c-e), the lines represent the rolling mean of the count of bubbles detected in each frame, with shading for the rolling standard deviation. All rolling quantities were computed using a window size equal to 20\% of the total number of frames in each dataset. Each measurement shows some fluctuations around a rolling mean that remains relatively steady throughout the measurement period.}
\label{fig:time}
\end{figure}

\subsection{Experimental protocol and statistically steady-state measurements of bubbles, drops, and particles} 

We present the systematic experimental protocol used to characterize drop production by bubble bursting in the statistically steady-state bubbling tank. Figure \ref{fig:time} illustrates the time series of the various measurements for a typical case (Broad-2~mm: broad-banded distribution of bubbles created by underwater turbulence which breaks the bubbles into many smaller sizes). 

Once the artificial seawater solution is in the tank, a lid is placed on top to seal the system. A constant stream of filtered, compressed air is then flowed through the air above the water surface to flush aerosols from the system.  The background aerosols escape through a single port in the lid, which is left open throughout the run to maintain equilibrium within the system. The flushing continues until the CPC component of the SMPS measures a particle count below 10 particles/cm$^3$, as shown in panel (b), which can be compared to an ambient particle count of $\sim$1000~particles/cm$^3$. This ensures that the sea salt aerosols we sample in the air are nearly all emitted by bubble bursting. The air through the top of the tank is then turned off, and the system is left to equilibrate for 40 minutes. CPC particle counts from the pre-run period are shown in panel (b) of figure \ref{fig:time}. We confirm that the particle count and relative humidity level (panel (a)) have reached a steady state before proceeding with the run. Relative humidity, air temperature, and water temperature are tracked throughout the run using a Thorlabs sensor (TSP01, TSP-TH) mounted to the tank lid. Values are comparable to realistic ocean conditions and remain steady throughout the run (temperatures vary by $\sim$1\textdegree{C} throughout the measurement period).

After the system reaches a steady state of particle counts, the SMPS, OPS, and holographic measurements begin simultaneously (shown in terms of the drop concentration in panel (b)). The SMPS and OPS sample continuously throughout the entire 50-minute run. Three independent holographic movies are recorded during the run, with approximately 15 minutes of saving time required between each measurement. For the bubbles (panels c, d, e), the large field-of-view surface measurement is made at the start of the run, followed by simultaneous measurements of bulk bubbles for both fields-of-view. Surface and bulk bubbles are recorded separately so that the back-lighting for each measurement does not interfere with the other. Finally, the small field-of-view surface measurement is made at the end of the run. Compared to the large surface measurement, the small surface images provide a significantly more zoomed-in view that requires more light than the large surface view, so the two measurements are made separately. Details of each measurement are given below and summarized in table \ref{tab:meas}.
Representative sample images for each bubble measurement are also shown in figure \ref{fig:time}(c-e) with detection overlaid. Some fluctuations are present in the number of bubbles and drops with time, but the magnitude of the fluctuations is fairly consistent throughout the run. Therefore, we consider the system to be running in a statistically steady state. 

\begin{table}
  \begin{center}
\def~{\hphantom{0}}
  \begin{tabular}{lcccc}
      Measurement  & Pixel Size (\textmu{m}) & Field of View (mm$^2$) & Depth of Field (mm) &  Recording Framerate (Hz) \\[3pt]
       bulk & 42 & 80.6 x 50.4 & 14\textendash26 & 1\\
       bulk small & 5.3  & 10.6 x 10.6 & 2 & 2\\
       surface & 10 & 53.3 x 46.1 & & 2 \\
       surface small & 1.6 & 3.28 x 3.28 & & 3\\
       drops - holography & 0.98 & 3.89 x 2.19 & 360 & 20\\
       drops - shadow & 29 & 59.4 x 59.4 & 120 & 5\\
  \end{tabular}
  \caption{Measurement details for imaging of bulk bubbles, surface bubbles, and drops (holography and shadow) in the present experiment. The shadow imaging drop measurement was performed for certain cases (Narrow - 2~mm, Narrow - 3~mm, Narrow - 250~\textmu{m}, and Broad - 2~mm) to allow for the measurement of drops very close to the surface. A range is specified for the depth of field of the bulk measurement resulting from the size-dependent depth-of-field correction, and this range corresponds to the depth of field of bubble radii in the range $R_b$ = [150, 3000]~\textmu{m} (see details in section 2.2.3 and figure 6 of \citet{methods}).} 
  \label{tab:meas}
  \end{center}
\end{table}

\subsubsection{Details of bulk bubbles}
\label{sec:bulk}
As introduced for the experimental setup in section \ref{setup}, two cameras measure the bulk bubbles simultaneously with two overlapping fields of view to enable measurement of bubbles with radii spanning $R_b$ = [30, 5000]~\textmu{m} (large field of view: Basler acA1920-40 with Zeiss Planar T50 lens, small field of view: Basler acA2040-90 with Nikon AF-Micro Nikkor 200 lens and 36~mm extension tube). The recording framerates of 1~Hz (large) and 2~Hz (small) are chosen so that a single bubble does not appear in multiple frames. Imaging details are provided in table \ref{tab:meas}.

Bulk bubble edge detection was performed using a Canny filter algorithm, which was done in multiple stages for the large field of view to enable successful detection of many sizes of bubbles.  An ellipse was fit to each detected region, and the bubble radius $R_b$ represents the volume-equivalent size of the bulk bubble, assuming axisymmetry around the minor axis of the ellipse. The void fraction $\phi_b$ is typically between 0.1-0.5\% for all cases studied, indicating a dilute region of bubbly flow. Through inspection of dedicated high-speed movies, we confirm that underwater collisions and coalescence in the bulk are very rare. The presence of salt limits the coalescence efficiency so that if bubbles do contact each other, they typically bounce off and do not coalesce.

The depth of field for all cases is calibrated to account for its dependence on the size of the bubble, using the method detailed in section 2.2.3 and figure 6 of \citet{methods}. For the Narrow - 2~mm and Narrow - 3~mm cases, the depth of field is additionally adjusted to account for the fact that the bulk bubbles are not evenly distributed throughout the depth (because they are rising from a square array of needles). This adjustment leads to a larger effective depth of field for the narrow-banded cases, which enables comparison to the broad-banded cases where the bubbles are distributed evenly throughout the volume by the underwater pumps. The average relative error on the bubble radius measurement is 6\% at the limits of the depth of field, see \citet{methods} (section 2.2.3 and figure 7).

\subsubsection{Details of surface bubbles}

At the surface, measurements of the bubbles are again made in two overlapping fields of view for the same range of bubble radii, $R_b$ = [30, 5000]~\textmu{m}. The camera for the large field of view (Basler a2A5328-15) captures clusters of bubbles, while the small field-of-view camera (Basler acA2040-90 with Infinity K2 DistaMax microscope lens) zooms in on the smallest bubbles that rise to the surface. The large and small surface measurements were made at 2~Hz and 3~Hz, respectively, and the frame independence of both measurements was checked by downsampling in time. Details are provided in table \ref{tab:meas}.

Surface bubbles are detected using the Hough Transform with custom filtering to remove duplicates and misdetection, which enabled detection of both clustered and isolated bubbles within the same frame using a single processing algorithm. The apparent radius of the detected bubbles from the surface measurement is then converted to the volume-equivalent radius $R_b$ by considering the static shape of the bubble at the surface, as described in \citet{neel2021}, based on the work of \citet{toba}.

\subsubsection{Details of drops and dry particles}

As the bubbles burst at the surface, they release drops into the air above the water surface. Three techniques are combined to measure ejected drops with radii spanning four orders of magnitude, in the range $r_d$ = [0.05, 500]~\textmu{m}. The first measurement is the digital inline holographic imaging setup (laser: CrystalLaser Nd:YLF QL527-200-L, camera: Phantom VEO4K-990-L, lens: Infinity K2 DistaMax, fully described in \citet{methods}). The holographic setup is positioned 5.5~cm above the water surface and captures the large sizes of drops, in the range $r_d$ = [10, 500]~\textmu{m}. The small field of view and framerate of 20~Hz are chosen so that each drop is counted only once, either on the upward or downward part of its trajectory.

Drops are then sucked out of the air above the tank and dried. The dry salt particles in the intermediate range of dry particle diameter, $D_p$ = [0.3, 10]~\textmu{m}, are counted and sized by the OPS. To improve the accuracy of the particle size measurement, an index of refraction of 1.5-0$i$ is specified for the sea salt and the 660~nm laser of the OPS \citep{shettlefenn}. The SMPS measures the smallest particles in the range $D_p$ = [0.02, 0.8]~\textmu{m}.  SMPS and OPS measurements have been used for atmospheric applications where the measured particle itself was of interest \citep{prather2013,wang2017}, as well as in other bubble bursting experiments where the OPS and SMPS measurements of salt particles are used to access the smallest sizes of drops that we are unable to measure using traditional optical methods \citep{sampath_aerosolization_2019,jiang2022, dubitsky_effects_2023}. 
To plot all three drop/aerosol measurements together, we convert the SMPS and OPS measurements of the dry particle diameter to drop radii by assuming conservation of salt mass in the drop and solving for the drop radius: $r_d = \frac{1}{2}D_p({\rho_{dry}}/\/\rho_s)^{1/3}$. From $\rho_s$ = 36~g/L and $\rho_{dry}$ = 2,056~g/L, we have $r_d \approx 2D_p$, as discussed in \citet{lewis_sea_2004}, which is a relation classically used to convert between the salt dry particle size and the liquid drop radius.

To quantify the background noise of aerosols remaining in the system, SMPS and OPS measurements were performed for a solution of DI water immediately before each artificial seawater case is run. The calculated background is then subtracted from the size distribution, smoothed using a Savitsky-Golay filter, and cut to the trusted measurement region. In order for all three measurements to be plotted together, the native units output by the SMPS are converted to the normalized concentration $N_d(r_d)$, such that integrating the distribution gives a number of particles per unit volume (see \citet{methods} for conversion details). 

We note that some drops above $r_d$ = 200~\textmu{m} may not be detected by the holographic measurement because they fall back to the surface before reaching the measurement region \citep{neel_velocity_2022}. To confirm the presence of these large drops, the drop measurements were supplemented with a shadow imaging measurement (for drops of radii $r_d >$ 150~\textmu{m}) for the Narrow-2~mm, Narrow-3~mm, Narrow-250~\textmu{m}, and Broad-2~mm cases. This measurement was added directly above the level of the water surface to quantify large drops that may not have reached the holographic measurement region (camera: Basler acA2040-90, telecentric lens: Opto-E TC4MHR096-C). Details of the shadow measurement are included in table \ref{tab:meas}, and the imaging setup is described for drop measurements in \citet{methods}.

\subsubsection{Typical bubble and drop size distributions} 
\label{sec:dist}

\begin{figure}
  \centerline{\includegraphics[scale=0.73]{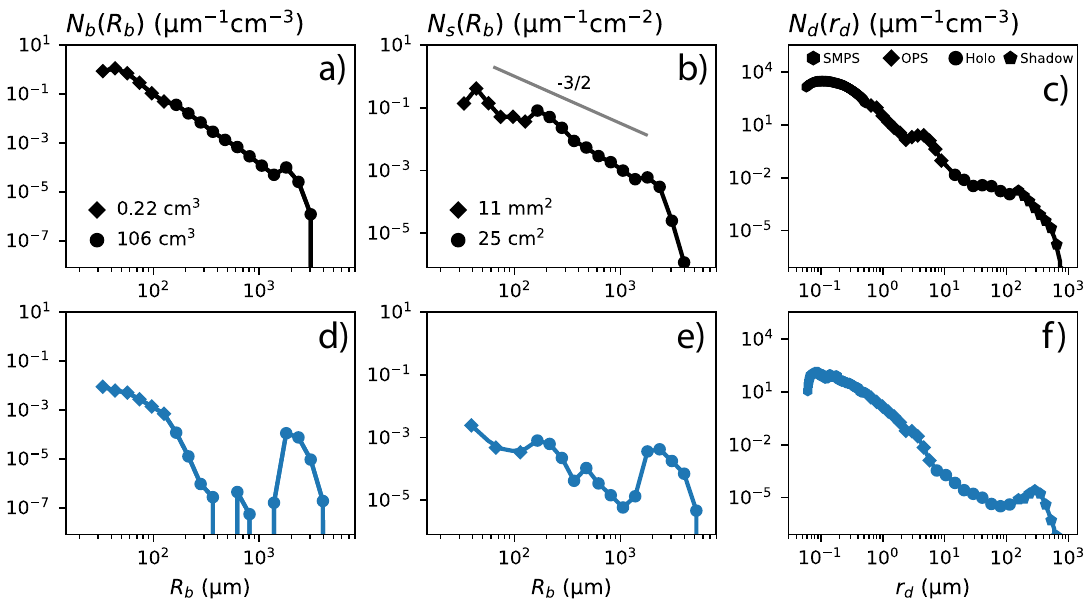}}
  \caption{Size distributions of bulk bubbles ($N_b(R_b)$, (a, d)), surface bubbles ($N_s(R_b)$, (b, e)), and drops ($N_d(r_d)$, (c, f)) for broad-banded (a-c) and narrow-banded (d-f) distributions of rising bubbles (see cases Broad-2~mm and Narrow-2~mm and in table \ref{tab:cases}). Different symbols indicate the different fields of view (bubble measurements) or the different measurement techniques (drop/particle measurements). The drop size distribution is shown in terms of the liquid drop radius, $r_d$. All distributions are interpolated over the logarithmically-spaced bins, with the interpolated distribution plotted over the data points.}
\label{fig:dist}
\end{figure}

The systematic measurements described above for a wide range of scales allow us to extract full size distributions for both drops and bubbles and probe different conditions in a systematic way (conditions summarized in table \ref{tab:cases}, with sample surface images in figure \ref{fig:surfims}). We illustrate the characterization of the size distribution for two representative cases in figure \ref{fig:dist}, the Broad-2~mm and Narrow-2~mm cases.

% sample data %

Using bubble sizes and counts extracted from the images through the methods described in the previous sections, we plot the size distributions of the detected bulk and surface bubbles, $N_b(R_b)$ and $N_s(R_b)$, shown in figure \ref{fig:dist} normalized by the bin size and by the measurement volume or area, respectively. For both bubble size distributions, data from the large and small field-of-view measurements are represented by different symbols and nicely overlap. Note that the bubble radius $R_b$ represents the volumetric bubble radius for both the bulk and surface measurements.

For the drop size distribution, all measurements are expressed as $N_d(r_d)$, the number of drops per unit bin size and unit measurement volume. Data from the SMPS, OPS, holographic imaging, and shadow imaging overlap well, and the combined measurements result in a single distribution that characterizes drops spanning 50~nm to 500~\textmu{m}. 

For the broad-banded case, we observe a continuum of bubble radii spanning $R_b$ = [30, 5000]~\textmu{m} in both the bulk and the surface (a, b). We see a corresponding continuum of drop radii spanning $r_d$ = [0.05, 500]~\textmu{m} (c). In the narrow-banded case, we see a narrow peak of bubble sizes around the injection bubble radius, $R_b$ = 2.3~mm (d, e). We refer to this case as narrow-banded, or nearly monodisperse, because the majority of bubbles are found in this peak around 2~mm. Some amount of smaller bubbles are also present for this narrow-banded case (formed as bubbles pinch off from the needles), and there are also smaller drops, shown in the drop size distribution below $r_d \sim$ 200~\textmu{m} (f). The bulk and surface distributions show similar trends in both conditions, suggesting limited coalescence (which is expected for artificial seawater). 

\subsection{Summary of the experimental analysis}

Having obtained the full size distributions of bubbles and drops, we now analyze the link between the different measurements to systematically quantify the attribution of drops to bursting surface bubbles. 
In section \ref{sec:bulksurf}, we link the measured bulk bubble size distribution (which integrates to a number per unit volume) with the surface bubble size distribution (which integrates to a number per unit area) using a simple flux argument involving the bubble rise velocity and the bubble lifetime at the surface. We then analyze the link between bursting bubbles and drops in section \ref{sec:surfdrop}. We consider a direct attribution of measured drops to measured bubbles in section \ref{sec:direct}, and in section \ref{section:forward} we predict a drop size distribution from the measured bubbles using existing scaling laws developed for individual bubble bursting. 

\section{Linking bulk and surface bubble size distributions} 
\label{sec:bulksurf}

\begin{figure}
  \centerline{\includegraphics[scale=.45]{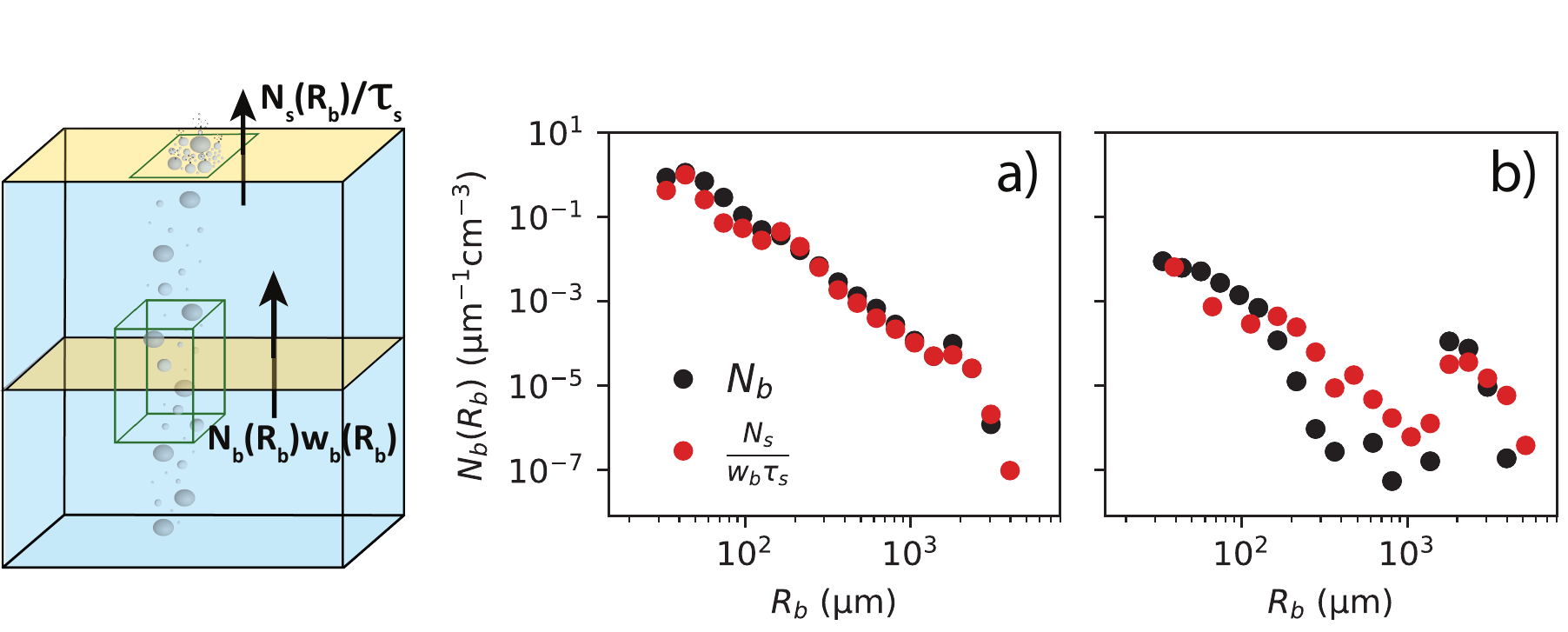}}
  \caption{Comparison of the measured bulk bubble size distribution $N_b(R_b)$ to the converted surface bubble size distribution, which is represented as a quantity normalized by a measurement volume following equation \ref{surftobulk}.  The comparison is shown for the medium bubble size cases of (a) broad-banded and (b) narrow-banded bulk bubble size distributions. The sketch illustrates the bulk measurement volume and surface measurement area (outlined in green, not to scale), as well as the fluxes through planes in the bulk ($N_b(R_b)w_b$) and at the surface ($N_s(R_b)/\tau_s$). Bubbles and drops not drawn to scale.}
\label{fig:flux}
\end{figure}

We now quantitatively connect the bulk bubble and surface bubble size distributions. The size distribution of bubbles bursting at the surface may evolve from the distribution seen in the bulk due to factors like coalescence, formation of child bubbles after bursting, and size-dependent time scales associated with the rising and bursting of the bubbles. For the artificial seawater solution used in this experiment, we demonstrate that the surface and bulk distributions can be related through simple assumptions used to equate a flux of bubbles through a plane in the bulk to a flux of bubbles at the surface, as illustrated by the sketch in figure \ref{fig:flux}.

The surface bubbles are measured over an area, so that $N_s(R_b)$ is a number per unit bin size per unit area, while the bulk bubbles are measured in a volume underwater, so that $N_b(R_b)$ is a number per unit bin size per unit volume. To link the two, we equate the flux of bubbles that rise through a plane underwater, $N_b(R_b)w_b(R_b)$, to a flux of bubbles arriving and bursting at the surface, $N_s(R_b)/\tau_s$, where $w_b(R_b)$ is the bubble rise velocity and $\tau_s$ the bubble lifetime at the surface. The link between the two distributions reads:
\begin{equation}
  N_b(R_b) = \frac{N_s(R_b)}{\tau_s\ w_b(R_b)}
  \label{surftobulk}
\end{equation}

The bubble rise velocity $w_b(R_b)$ is a function of the bubble size, and we use the semi-empirical rise velocity parameterization for contaminated water presented in \citet{clift_bubbles_1978}, p. 172. The timescale required for the flux of surface bubbles is the bubble lifetime $\tau_s$. Significant uncertainty remains surrounding the dependence of bubble lifetime on the bubble size and properties of the solution (see table 1 in \citet{poulain_ageing_2018}), so for simplicity we consider a reasonable estimate of a constant bubble lifetime of order \textit{O}(1) second (we assume $\tau_s = 0.6$~s). Equation \ref{surftobulk} effectively neglects the effect of coalescence at the surface, which is generally an appropriate simplification given the seawater salinity of the solution \citep[effect of salts on bubble coalescence reviewed in][]{firouzi_saltrev}. 

The comparison between the bulk size distribution and converted surface distribution (calculated using equation \ref{surftobulk}) are shown in figure \ref{fig:flux} for the two typical cases presented in the previous section (figure \ref{fig:dist}, Broad-2~mm and Narrow-2~mm). We see good agreement in the overall trends and magnitudes of the measured bulk distributions and the converted surface distributions for both the (a) broad-banded case with turbulence present and (b) narrow-banded distributions with no turbulent flow. This indicates that, for this configuration, 1) coalescence is not a dominant process to account for, and 2) a constant $\tau_s$ is a reasonable approximation. 

Some words of caution are required when presenting the flux-based comparison. We have chosen one particular rise velocity parameterization and a constant bubble lifetime to evaluate the flux argument. Other relations for the rise velocity were also tested, including the parameterization from \citet{woolf1991} which gives a similar result aside from a slightly different velocity for the smallest bubbles. We could also have considered the effect of the background turbulence on the rise velocity, which systematically slows down the rise velocity of the bubbles \citep{ruth_effect_2021,liu24} (we note that a systematically lower rise velocity could be compensated for by a slightly higher value of the bubble lifetime, within the uncertainties of experimentally-measured bubble lifetime). Note that by integrating $\int_A(\int_{R_{b1}}^{R_{b2}} N_b(R_b)\frac{4}{3}\pi R_b^3w_{b}(R_b)\,dR_b)\,dA\,$ over the total range of bubbles sizes [$R_{b1}$, $R_{b2}$] and the area $A$ of the plane through which the bubbles rise, we do obtain air flowrates that are consistent with the total prescribed air flowrate given in table \ref{tab:cases} (and controlled primarily by the largest bubbles). By assuming equation \ref{surftobulk} and a rise velocity function, we could also have extracted an effective bubble lifetime function. While uncertainty remains in both the chosen rise velocity and lifetime, we find that we can relate the bulk and surface bubble size distributions well for all the cases presented in the paper, so that we can use the bulk and surface measurements interchangeably by applying equation \ref{surftobulk}. 

% now practical reason:
In the following analysis, we will use an effective size distribution for bursting bubbles (using both surface and bulk data) to practically relate the bubbles to the drop size distribution, which is measured per unit volume. A direct link between bursting bubbles and emitted drops in terms of fluxes would require measurements of the drop velocity (not available here). To come up with this effective bubble size distribution, the surface bubble distribution is first converted to a quantity of bursting bubbles normalized by a volume, following equation \ref{surftobulk}. For bubbles below $R_b$ = 150~\textmu{m}, we use data from the bulk small field-of-view measurement to construct a complete size distribution of bubbles spanning $R_b$ = [30, 5000]~\textmu{m}. Throughout the rest of the paper, the term $N_b(R_b)$ is used to represent this size distribution of bursting bubbles from the combined measurements, with the exception of the Narrow-250~\textmu{m} case, which uses only the converted surface data due to imaging/detection difficulties for bubbles in the denser bulk plume.

\section{Linking drops and bursting bubbles} 
\label{sec:surfdrop}
We can now analyze the link between the surface bubbles and ejected drops in the collective bursting configuration by examining how the drop size distribution changes as the bubble size distribution is intentionally varied through the different cases. 

\subsection{Direct attribution between measured bubbles and drops}
\label{sec:direct}

\begin{figure}
  \centerline{\includegraphics[scale=0.73]{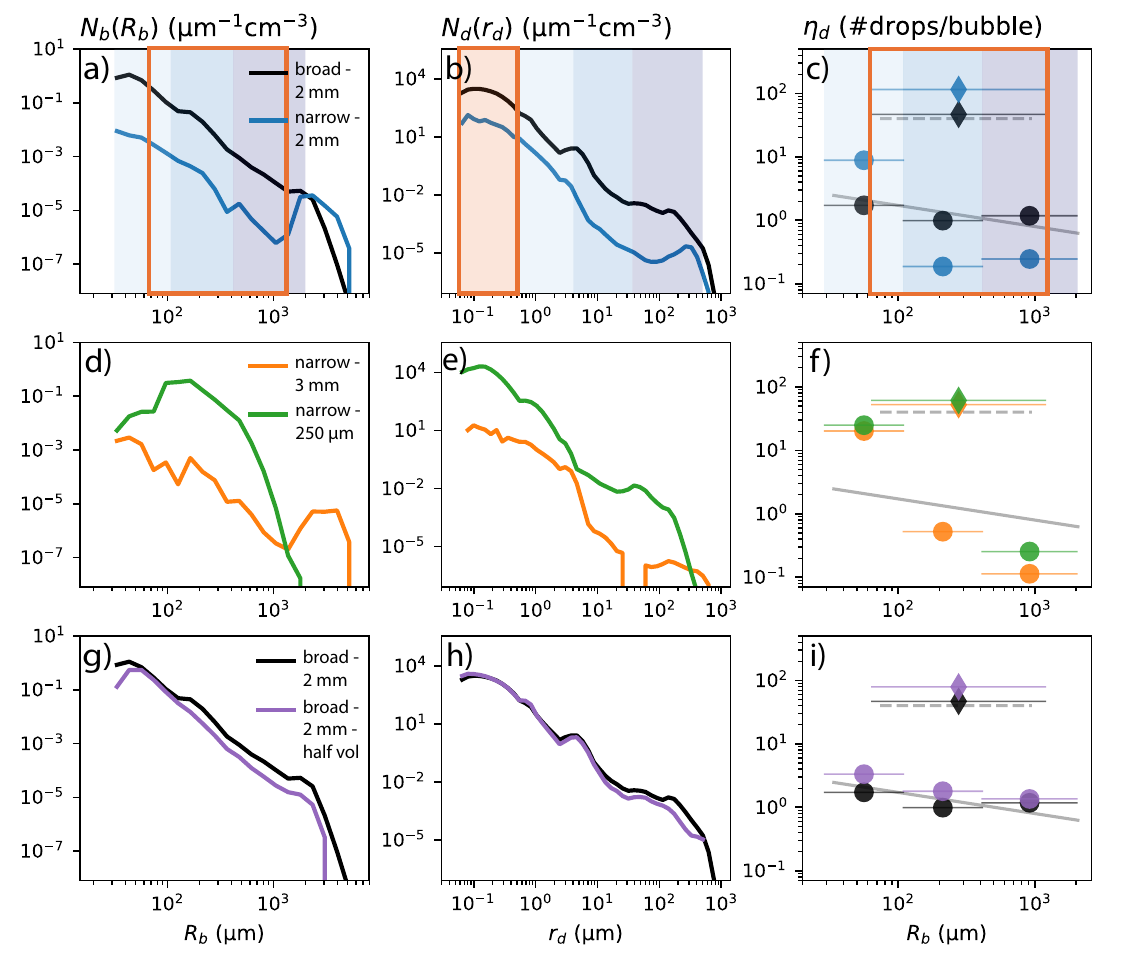}}
  \caption{Size distributions of bursting bubbles ($N_b(R_b)$, (a, d, g)) and drops ($N_d(r_d)$, (b, e, h), in terms of the liquid drop radius, $r_d$) for various case comparisons. The bubble size distributions presented vary significantly, including broad-banded cases and three narrow-banded cases peaking at radii of 2~mm, 3~mm, and 250~\textmu{m}, with a corresponding variety of drop size distributions. To calculate the drop production efficiency presented in the rightmost column (c, f, i), the bubble and drop size distributions are first integrated in the shaded or outlined size buckets, where the link between the bubble and drop sizes approximates the link suggested by scaling laws for individual bubble bursting. The resulting number of drops in each size range is then divided by the number of bubbles in the corresponding shaded size range (outlined for film flap range) to obtain an average number of drops produced per bubble in each bubble size bucket (production efficiency $\eta_d$, shown as circles within the jet drop size range corresponding to the blue shaded regions and diamonds for the assumed film flap drop size range corresponding to the orange boxes). Shaded gray lines show the number of drops per bubble predicted by the individual bubble bursting scalings for jet (solid line) and film flapping (dotted line) production mechanisms (tested in section \ref{section:forward}, see table \ref{tab:paramsfull} for details). Cases shown are (a-c): Narrow-2~mm vs. Broad-2~mm, (d-f): Narrow-3~mm vs. Narrow-250~\textmu{m}, and (g-i): Broad-2~mm vs. Broad-2~mm-half volume. Further details of each case are found in table \ref{tab:cases}, and the uncertainty associated with the production efficiency values shown in (c, f, i) is discussed in appendix \ref{appB}.}
\label{fig:invattr}
\end{figure}

We first show the drop size distributions resulting from various bubble size distributions and discuss how different sizes of drops may be attributed to different sizes of bursting bubbles. Figure \ref{fig:invattr} shows the size distributions of bursting bubbles ($N_b(R_b)$, (a, d, g)) and drops ($N_d(r_d)$, (b, e, h)) as well as the ratio of the number of drops produced per bubble ($\eta_d$, (c, f, i)) in chosen size ranges.

The first row (a-c) of figure \ref{fig:invattr} shows the two sample cases described in section \ref{sec:dist} (figures \ref{fig:dist} and \ref{fig:flux}, Narrow and Broad 2~mm injection cases). Both cases use the same needles and flow rate, but in the broad case turbulent flow is created underwater. As described above, turbulence induces bubble breakup and creates a broad-banded bubble size distribution with a continuum of bubbles spanning radii of 30~\textmu{m} to almost 5~mm. Without turbulence, the distribution has a peak around $R_b$ = 2~mm with a lower number of smaller bubbles. The two cases (broad- and narrow-banded) have a relatively comparable number of bubbles around 2~mm, near the needle injection size. For the submillimetric bubbles, there is a large difference of about two orders of magnitude in number of bubbles (because without turbulence, the only small bubbles present are formed by pinch-off at the needles, see section \ref{sec:start}). Moving to the drop size distribution, we observe a corresponding difference of similar magnitude between the drop size distributions for nearly the entire range of sizes measured in the experiment, until the number of drops becomes more comparable between cases for $r_d >$ 400~\textmu{m}. This suggests that the drops we measure below radius 400~\textmu{m} are produced by bubbles with radii smaller than $\sim$2~mm (shaded in color).

Thus, we focus on the submillimetric portion of the bubble size distribution (shaded in color) and consider the link between the measured submillimetric bubbles and produced drops. We first test a direct attribution of drops to bubbles, dividing the number of measured drops by the number of measured bubbles in discrete size bins (shaded/outlined with the corresponding colors).

Size buckets are chosen to approximate the individual bubble size scaling laws for jet and film flapping production mechanisms (discussed in more detail in section \ref{section:forward}, and \citet{berny2021,jiang2022,deike_mechanistic_2022}). The corresponding bubble and drop ranges are shown in shaded/outlined colors in figure \ref{fig:invattr}, where the overlap in bubble sizes represents the fact that bubbles can produce both jet and film flap drops in parallel processes. For the film flap scaling, we test the link between bubbles of radii $R_b$ = [70, 1200]~\textmu{m} and drops of radii $r_d$ $<$ 0.5~\textmu{m} (outlined/shaded in orange). For the jet drop scaling, we consider the ability of bubbles in the range $R_b$ = [30, 2000]~\textmu{m} to produce drops of $r_d$ $>$ 0.5~\textmu{m} (shaded in blue). The total regions tested for jet drop production are approximately evenly divided into 3 smaller log-spaced size buckets for both bubbles and drops, and the buckets are linked in size order. 

The distributions are then integrated within each assigned size bucket to calculate a total number of bubbles or drops (per measurement volume) within that size range, given by $\rho_b = \int_{R_{b1}}^{R_{b2}} N_b(R_b) \,dR_b$ for bubbles within the size range $[R_{b1},R_{b2}]$, and $\rho_d = \int_{r_{d1}}^{r_{d2}} N_d(r_d) \,dr_d$ for drops in the size range $[r_{d1},r_{d2}]$. By dividing the number of drops in a size bucket by the number of bubbles in the corresponding size bucket, we calculate a drop production efficiency (the number of drops produced per bubble) for each pair of size buckets, defined as $\eta_d = \rho_d/\/\rho_b$. 

Resulting drop production efficiencies are shown in the right column of figure \ref{fig:invattr} (c, f, i), calculated from the shaded ranges (jet drop, circles) or outlined ranges (film flap drop, diamonds) of bubble and drop size buckets for the two cases. Note that the values of the efficiencies would shift somewhat if we chose to split the drop size distribution at a value other than $r_d$ = 0.5~\textmu{m} when testing the attribution to the assumed bubble size ranges. The cutoff radius at $r_d$ = 0.5~\textmu{m} was chosen because it falls between the largest predicted film drop and smallest predicted jet drop for the range of bubble sizes present in the experiment. 

For the broad-banded case, we observe $\sim$50 drops/bubble in the assumed film flap range (c, black diamond) and $\sim$1.0\textendash1.8 drops/bubble across the jet drop size buckets (c, black circles). For the narrow-banded case, we observe a high efficiency of around 100 drops/bubble in the film flap size range (c, blue diamond), while the small jet drop efficiency is around 9 and the larger jet drop efficiency is below unity (c, blue circles).

These numbers can be compared to the efficiencies discussed in the literature. The light gray lines in the same plot represent the drop number given by the individual bubble bursting scaling laws used throughout the paper for jet drops (solid, between 0.6\textendash3 drops produced per bubble, \citet{wang2017}) and film flap drops (dotted, 40 drops per bubble, \citet{jiang2022}). Detailed equations for the scalings shown here, based on past work in the literature, are provided in table \ref{tab:paramsfull} and described in section \ref{section:forward}. 

Comparing the calculated efficiencies for the two sample cases to the approximate individual bursting scaling laws, we see reasonable agreement for the broad-banded case. For the nearly monodisperse case, significant differences are observed, and various interpretations can be proposed. The higher efficiency of film flap drops could be related to the proposed bubble pinch-off effect/small aerosol formation in the bulk discussed in the recent work from \citet{jiang_abyss_2024}, with this effect potentially smeared out or not occurring in the presence of turbulence (note that for the Narrow-2~mm case, high-speed movies confirm collisions at the needle tip/during bubble pinch-off as discussed in \citet{jiang_abyss_2024}. These collisions at the needle tips are not observed in the Narrow-3~mm case). The higher jet drop efficiency for the small bubbles, which closely match direct numerical simulations for similar conditions \citep{berny2021}, could be due to the fact that either low surface density of bubbles or a calmer free surface are necessary to obtain such efficiency. Larger jet drop low production efficiency might be related to a lack of statistical convergence in measuring these drops with the holographic system. Sources of uncertainty involved in the calculation of the production efficiency values presented in figure \ref{fig:invattr} are discussed in appendix \ref{appB}. 

To further probe the link between collective bursting bubbles and ejected drops, we perform the same direct attribution analysis for several initial bubble configurations and analyze the corresponding change in the drop size distribution and associated uncertainties.  We look next at the middle row of figure \ref{fig:invattr} (d-f) to analyze how injecting bubbles larger or smaller than 2~mm, with a nearly monodisperse size distribution, affects the drop production. For the injection of larger bubbles (d), the peak of the size distribution is closer to a radius of $\sim$3.5~mm (compared to the previous case of $\sim$2.3~mm bubbles). The small bubble case features a wider peak of primarily submillimetric bubbles centered near 250~\textmu{m} (see details in table \ref{tab:cases} and surface images in figure \ref{fig:surfims}).

From the size distributions, it is immediately evident that many more submicron droplets are produced (e) when more small submillimetric bubbles are present (d). We observe an approximately \textcolor{blue}{3} order-of-magnitude difference between the size distributions of the small bubble case and the large bubble case, where very few submicron drops are produced. As shown by the large bubble case, when we measure very few submillimetric bubbles, we measure very few submicron drops. This demonstrates that bubbles predominantly larger than 2~mm do not produce a significant amount of submicron aerosols in this configuration. 

Comparing to individual bursting scalings, we see a reasonable efficiency of $\sim$60 drops/bubble for both cases for the proposed film flap range (f, green and orange diamonds). In the two largest jet drop bins, the efficiencies for bubbles are low (f, green and orange circles; efficiencies below 0.1 associated with very small number of drops are not shown on the plot) with similar values to those discussed above for the 2~mm narrow-banded case (c, blue circles). For the smallest jet drop bin, the efficiencies are high, at around 20 drops/bubble. We note that the overall efficiencies are consistent across all the cases despite the large differences in the bubble size distributions (see the plot of number of drops versus number of bubbles in appendix \ref{appA}), and it is only once we assume a link between the bubbles sizes and drop sizes for this attribution that differences in efficiencies appear for the narrow-banded cases compared to the broad-banded cases. The different efficiency values found in the narrow-banded cases imply that some subtleties of the bubble-drop link may be blurred across neighboring size buckets for the continuum of bubble and drop sizes present in the broad-banded cases, which seem to match the single bubble scaling laws well.

The lack of large drops for two of the narrow-banded cases (peaking at radii of 2~mm and 3~mm) can be explained by either missing drops due to the small field of view or by a physical collective effect where neighbors may somehow limit production efficiency in these nearly monodisperse conditions. 

To take a step in investigating a potential collective effect that may modulate the bubble to drop size link or affect overall production efficiencies, we look at the broad-banded cases and compare cases with 32 versus 16 bubbling needles (``Broad-2~mm" and ``Broad-2~mm-half volume" \textendash\ the cases have the same flowrate per needle, so we effectively halve the injected volume of bubbles). Aside from decreasing the injected volume, decreasing the number of needles allows the bubbles to arrive at the surface spaced apart from each other, so that most bubbles burst individually instead of clustered, as visualized in figure \ref{fig:surfims}(f). Looking at the size distributions for these two cases, shown in the bottom row of figure \ref{fig:invattr} (g-i), we observe that a smaller number of bursting bubbles (g) results in a drop size distribution that is very similar to that of the clustered case, especially for the submicron drops (h). Looking at the efficiencies (i), the 16 needle case has a higher efficiency than the 32 needle case across all bubble sizes, indicating that a collective effect could potentially be limiting the drop production efficiency. 

Through the attribution of measured drops to measured bubbles analyzed in this section, we have observed that the broad-banded cases show an overall agreement between our calculated bursting efficiencies and those found from the single bubble scaling laws for film flap and jet drop production, with a possible reduction of efficiency for the collective bursting (g-i). The narrow-banded cases demonstrate that submicron drops are produced by bubbles with radii smaller than $\sim$1~mm (d-f), and that nearly monodisperse clusters may display a reduced efficiency in jetting for bubbles with radii above $\sim$100~\textmu{m} and an increased efficiency for the smaller bubbles (c, f).

\subsection{Using individual bubble scaling laws to predict drops produced by measured bubbles} 
\label{section:forward}

Now we investigate the distributions of drops predicted by the individual bubble bursting scaling laws \textendash\ which provide number and size of drops for a given bubble \textendash\ combined with the measured bubble size distributions, and we then compare the predicted drop size distributions to our measured drop size distributions. We use the results to further evaluate how well the individual bursting relations describe collective bursting results across all cases. 

To predict the drops produced by the collection of bursting bubbles, we consider the approach from \citet{lv2012}, initially proposed for production of film drops via a centrifuge mechanism and extended to jet drops \citep{berny2021} and film flap drops \citep{deike_mechanistic_2022}. 
For a specific drop production process, the total drop size distribution $N_d(r_d)$ produced by an ensemble of bubbles is obtained by integrating individual bursting scalings over the size distribution of bursting bubbles $N_b(R_b)$ in the size range [$R_{b1}$, $R_{b2}$]:

\begin{equation}
  N_d(r_d) = \int_{R_{b1}}^{R_{b2}} \frac{N_b(R_b)n_{d}(R_b)}{\langle r_d \rangle(R_b)} p\left(\frac{r_d}{\langle r_d \rangle}, R_b\right) \,dR_b\, ,
  \label{globaldrop}
\end{equation}\\
where $N_b(R_b)$ is the measured bubble bursting size distribution, $\langle r_d \rangle(R_b)$ is the mean drop size produced by a bursting bubble of size $R_b$, $n_{d}(R_b)$ is the number of drop produced by a bubble bursting of size $R_b$, and $p(r_d/\langle r_d \rangle, R_b)$ is the probability density function of drop sizes produced by a bubble of size $R_b$, assumed to be a Gamma distribution of known order \citep{lv2012,berny2021,deike_mechanistic_2022}. Sensitivity of the resulting drop size distributions to the order of the Gamma distribution and to the chosen radius bounds ($R_{b1}$ and $R_{b2}$) are discussed in appendix \ref{appB}.

We perform the integration, using the measured bubble size distribution, for each individual drop production mechanism, as summarized in table \ref{tab:paramsfull}. Scalings for the number and size of drops produced by bursting of single bubbles have been developed for three main modes of drop production: film centrifuge, film flap, and jet drop production. The integration for each mechanism over many bubble sizes is similar to the analysis performed in \citet{deike_mechanistic_2022} (using assumed/modeled bubble size distributions) and \citet{neel_surf_2022} (measuring only larger scales of drops and bubbles), as well as for individual mechanisms in \citet{lv2012} (film centrifuge drops, assumed/modeled bubble size distribution) \citet{blancorodriguez_sea_2020} (jet drops), and \citet{ganan-calvo_ocean_2023} (extended jet drops using assumed/modeled bubble size distributions). 

\begin{figure}[!h]
  \centerline{\includegraphics[scale=0.9]{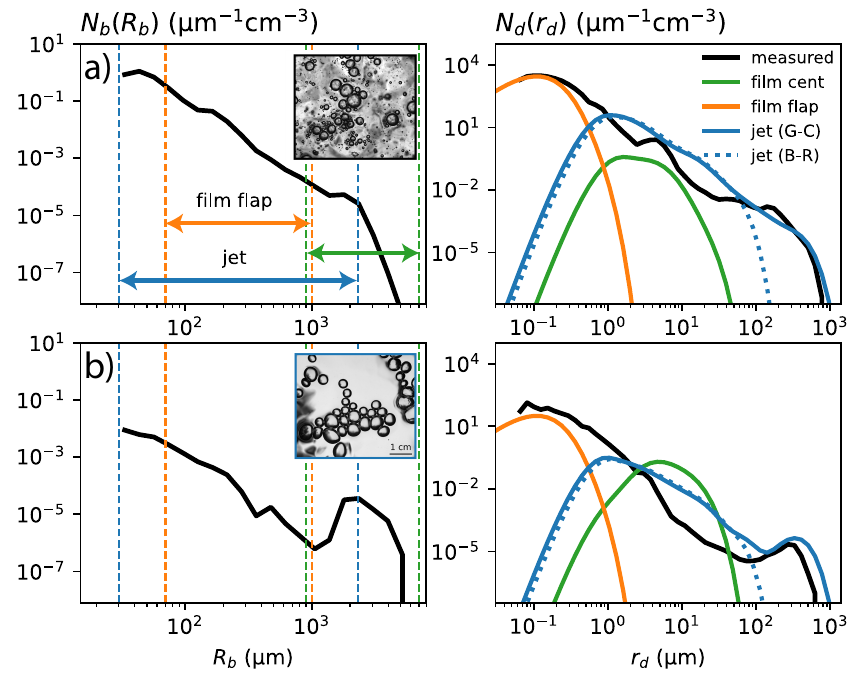}}
  \caption{Measured bubble ($N_b(R_b)$) and liquid drop ($N_d(r_d)$) size distributions for the 2~mm injected bubble cases with (a) broad-banded and (b) narrow-banded distributions. Overlaid colored lines represent the predicted drop size distribution from each production mechanism for individual bubble bursting (see details in table \ref{tab:paramsfull}), integrated over many bubble sizes using the measured bubble size distribution. Bubble size ranges used in the integration for each mechanism are labeled in the same color on the measured bubble size distributions. Note that the jet (B-R) formulation is only applied in the range $R_b$ = [30,500]~\textmu{m}.}
\label{fig:attr4lines}
\end{figure}

Figure \ref{fig:attr4lines} shows the result of the integration \ref{globaldrop} for each production mechanism, compared to the measured drop size distribution for the 2~mm bubble size cases (broad-banded and narrow-banded). The scalings and coefficients used for each drop production mechanism are given in table \ref{tab:paramsfull} and depicted graphically in figure \ref{fig:invattr}.

\begin{table}
  \begin{center}
\def~{\hphantom{0}}
  \begin{tabular}{lcccc}
      Mechanism  & [$R_{b1}$, $R_{b2}]$ (\textmu{m}) & $\langle r_{d} \rangle(R_b)$ (\textmu{m}) & $n_d(R_b)$ & Order of $\Gamma$ \\[1pt]
        &  &  & & distribution \\[4pt]
       film flap & [70, 1000] & $0.40\ \left(\frac{R_b}{l_c}\right)^{1/3}$ & 40 & 4\\ [2pt] 
       - & - & [0.1, 0.3]~\textmu{m} & - & -\\[5pt] 
       jet (G-C) & [30, 2300] & $0.60\ l_{\mu}\ \left( \sqrt{La}\left(\sqrt{\frac{La}{La_*}} - 1\right)\right)^{5/4}$ &
  $30 \left(\frac{R_b}{l_{\mu}}\right)^{-1/3}$ & 11\\[2pt] 
       - & - & [1, 400]~\textmu{m} &
   [3, 0.6] & -\\[5pt]
       jet (B-R) & [30, 500] & $0.20\ R_b \left(1 - \left({\frac{0.033^2 R_b} {l_\mu}}\right)^{-1/4}\right)$ & $30 \left(\frac{R_b}{l_{\mu}}\right)^{-1/3}$ & 11\\[5pt] 
       film centrifuge & [900, 10 000] & $0.25\ R_b^{3/8}\ h_b^{5/8}$ & $0.040 \left(\frac{R_b}{l_c}\right)^{2} \left(\frac{R_b}{h_b}\right)^{7/8}$& 11\\
        - & - & [1, 50]~\textmu{m} & [10, 120] & -\\[2pt]
  \end{tabular}
  \caption{Equations and parameters used to calculate the drop size distribution $N_d(r_d)$ from input bubble size distribution $N_b(R_b)$, following (with adaptation) the scaling laws developed for individual bubble bursting. Equations used for the mean size $\langle r_{d} \rangle(R_b)$ and number $n_d(R_b)$ of emitted drops are shown, as well as the radius bounds [$R_{b1}$, $R_{b2}$] and the order of the Gamma distribution (assumed to represent the distribution of drop sizes $p(r_d/\langle r_d \rangle, R_b)$) required to compute $N_d(r_d)$ as described in equation \ref{globaldrop}. Film flap: Size scaling based on the data from \citet{jiang2022}, discussed in \citet{deike_mechanistic_2022}. The drop number is taken as a single value of 40 drops/bubble, based on \citet{jiang2022}. Jet (G-C)/Jet (B-R): Uses the formulation for the first drop size given by \citet{ganan-calvo_revision_2017} and \citet{blancorodriguez_sea_2020}, respectively. The drop number scaling from \citet{berny2021} is used with a modified prefactor suggested by \citet{wang2017}. The Laplace number $La = R_b/l_{\mu}$ controls jet drop ejection, comparing the radius of the bursting bubble, $R_b$, to the visco-capillary length, $l_{\mu} = \mu^2/\/\rho\gamma$, using the values for physical parameters given in section \ref{setup}. $La_* = 550$ is taken for the critical Laplace number for jet drop formation in the G-C formulation \citep{walls_jet_2015, berny_role_2020}. Note that the B-R expression provided here is valid for $Bo = \rho gR_b^2/\gamma \leq 0.05$ and $La >$ 1111, which corresponds here to bubbles in the radius range $R_b$ = [20, 500]~\textmu{m}, so this formulation is applied only to our measured bubbles of radii smaller than 500~\textmu{m}. The G-C scaling is applied to bubbles of radii larger than 30~\textmu{m} and would diverge for smaller radii between 10 \textendash\ 20~\textmu{m}. Film centrifuge: Scalings developed by \citet{lv2012}, with size and number of drops determined by the bubble size and film thickness $h_b$, where $h_b$ scales with {${R_b}^2$/\/$l_c$}  and spans [0.2, 30]~\textmu{m} for bubbles with $R_b > 0.4l_c$ (capillary length $l_c = \sqrt{\gamma/\/\rho g} = 2.3$~mm for the artificial seawater solution). Sensitivity of the predicted drop size distributions to different choices of $R_{b1}$, $R_{b2}$, and the order of the Gamma distribution are discussed in appendix \ref{appB}.}
  \label{tab:paramsfull}
  \end{center}
\end{table}

From figure \ref{fig:attr4lines}, we can discuss how well scaling laws for each drop production mechanism describe the experimental drop size distribution, beginning with the broad-banded case presented in the top row.

%jet
We start with jet drop production, where drops are created by fragmentation of the jet formed by the focusing of capillary waves into the collapsed bubble cavity. For bubbles of sizes between 30~\textmu{m} and the capillary length ($l_c$ $\sim$ 2.3~mm for the artificial seawater solution), which are able to produce jet drops \citep{berny_role_2020,walls_jet_2015}, we apply the theoretical formulation developed by \citet{ganan-calvo_revision_2017} for the resulting drop size. Because there is no theoretical prediction for the number of jet drops produced by the bursting bubble, we consider the scaling proposed by \citet{berny2021} for the number of drops (coherent with experimental data from \citet{ghabache_physics_2014, ghabache_size_2016, spiel_jet_97}) with a modified prefactor. The modified prefactor means that bubbles of radii $R_b$ = [30, 2300]~\textmu{m} are predicted to produce $n_d$ = [3, 0.6] jet drops per bubble, which is lower than the $n_d$ = [11, 3] drops from the \citet{berny2021} fit to various data for individual bubble bursting, but is in agreement with the collective bursting experiment from \citet{wang2017} (see figure 1 from their paper).

For the broad-banded case, shown in the top row of figure \ref{fig:attr4lines} (a; bubbles on the left, drops on the right), the jet drop scaling laws (blue lines) match the measured drop size distribution (in black) very well for drops of radii approximately 1~\textmu{m} to 500~\textmu{m}. Because the magnitude of the distributions lines up well using the 0.6\textendash3 jet drops produced per bubble across the radius range (where smaller bubbles produce more drops), the average number of jet drops produced by many collective bursting events is likely lower than the 3\textendash11 drops that a single bubble is physically able to produce (across the same radius range). Each bursting event is not necessarily as efficient as every other event. In the collective bursting setup, some bubbles may be disturbed by the rising or bursting of their neighbors or by water fluctuations that could affect the cavity collapse and jet formation, which would reduce the overall drop production efficiency.
The jet drop size formulation developed in \citet{blancorodriguez_sea_2020}, based on capillary wave selection, was also tested (blue dotted line). Since it yields very similar results over the range of bubble radii we measure where the scaling is valid (as the drop sizes as a function of bubble sizes were derived using the same experimental and numerical data in both formulations), we use the equation from \citet{ganan-calvo_revision_2017} for the rest of the analysis.

%film flap
Next we consider the mode of film drop production proposed in \citet{jiang2022}, where they report submicron drops produced by bursting bubbles in the approximate radius range $R_b$ = [70, 1000]~\textmu{m}, which they attribute to a film flapping mechanism. As shown in figure \ref{fig:attr4lines}, the size scaling and order of magnitude associated with the film flap drops (orange) match well with the measured distribution (black) for the broad-banded case. 

%film centrifuge
Finally, for film drop production through a centrifuge mechanism, \citet{lv2012} describe the fragmentation of ligaments formed as the cap of a bubble with radius $R_b > 0.4l_c$ retracts. We observe that the predicted distribution (green) does not describe the data well, falling below the measured distribution for the broad-banded case. The film centrifuge scaling leads to significantly fewer drops than the jet drop counterpart (in the size range where they both mechanisms are applicable) due to the fact that there are fewer larger bubbles (above the capillary length) where the centrifuge mechanism is effective. Note that the film centrifuge scaling does not predict submicron drops.

Looking now at the narrow-banded case presented in figure \ref{fig:invattr}(b), we observe that the film flap results again match the peak and magnitude relatively well (orange line), only slightly below the experimental data (black line). The jet scalings (blue) approximate the correct drop magnitude at their peak, around 1~\textmu{m}, but overpredict some larger drops. While this overprediction could be related to limits in measuring larger drops in the experiment, we also note that if a truncated Gamma distribution is used, such that $p(r_d/\langle r_d \rangle, R_b)$ = 0 for bubbles of radii $R_b >$ 2~mm, the resulting predicted drop size distribution describes the large drops well (shown in appendix \ref{appB} for the Broad-2~mm and Narrow-2~mm cases). The film centrifuge mechanism again does not describe the data, this time overpredicting the number of drops. This significant overprediction may be related to fact that the centrifuge mechanism can frequently eject drops horizontally, reducing their measured efficiency as these drops likely fall back to the water without being transported upwards into the measurement region \citep{neel_velocity_2022,shaw_film_2024}.

From the two sample cases, we find that the experimental drop size distributions can be predicted well from knowledge of the experimental bubble size distribution using scaling laws for film flap drop and jet drop production for individual bubble bursting, with some differences between data and model for jet drops above $r_d$ = 10~\textmu{m}. The film centrifuge mechanism does not appear to be statistically relevant and is ignored in the rest of the analysis. In the next section, we will explore how well this framework describes spray generation for the other bubble size distributions.

\begin{figure}
  \centerline{\includegraphics[scale=.75]{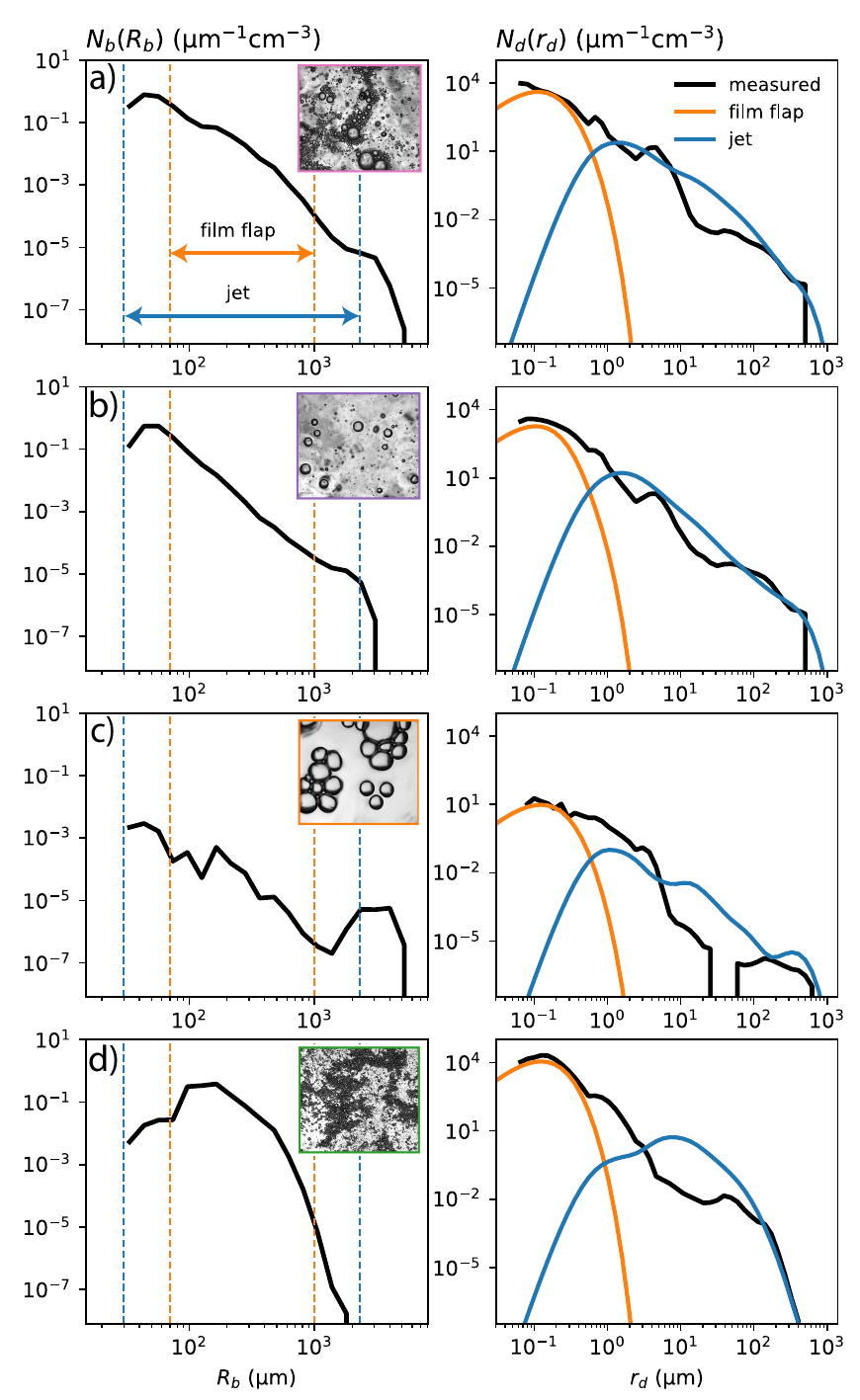}}
  \caption{Bursting bubble ($N_b(R_b)$) and liquid drop ($N_d(r_d)$) size distributions for the cases: (a) Broad-3~mm, (b) Broad-2~mm-half volume, (c) Narrow-3~mm, and (d) Narrow-250~\textmu{m}, with representative surface images inlaid for each case. The predicted drop size distributions from individual bubble bursting scaling laws are plotted in color for film flapping (orange) and jet (blue) drop production (see details in table \ref{tab:paramsfull}), with the bounds of the associated portions of the bubble size distribution shown with dashed lines of the same color.}
\label{fig:fwd}
\end{figure}
\par

\subsection{Jet and film drop individual bubble bursting scaling laws for all cases}
\label{sec:fwd2}
Figure \ref{fig:fwd} compares the measured and predicted drop size distributions for the remaining cases, using the same set of parameters described in table \ref{tab:paramsfull}. As introduced in the previous section, the individual bubble scaling laws match the data very well for the broad-banded cases, shown in figure \ref{fig:fwd}, row (a; bubbles on the left, drops on the right) for bubbles of original injection size 3~mm, which are again characterized by a continuum of bubble sizes after breakup in turbulence. The magnitude of the distributions again match well with the data (black) using the 0.6\textendash3 drop per bubble assumption for the jet drops (blue), as discussed in the previous section, and the film flap (orange) magnitude also lines up well with the data. In this case, the framework works well across all scales with the two modes of production \textendash\ with the jet drop mechanism dominating drop production above 1~\textmu{m} and the proposed film flap mechanism producing submicron drops.

Row (b) shows the distributions for the case with 16 bubbling needles, where bubbles are spaced out on the surface and burst individually. While the jet drop scaling still matches well, the film flap prediction seems to miss some measured drops. The slight underprediction for this case \textendash\ compared to the better match for submicron drops in (a) \textendash\ may indicate a higher efficiency for the individual bursting case due to a potential collective effect limiting drop production. 

For the cases with narrower distributions, shown in rows (c, d) of the same figure, the measured and predicted distributions match well in the size range of film flap drops but are less coherent for the region described by jet drop production. In the large bubble case (row (c)), again fewer large jet drops are measured than predicted. For the small bubbles, the prediction overestimates drops around radius $r_d$ = 10~\textmu{m}. As discussed in section \ref{sec:direct} for the direct attribution, the narrow-banded cases reveal that the assumed size link between bubbles and drops is not entirely accurate in this collective bursting setup when large aggregate of bubbles of nearly the same size are present. 
Indeed, it seems that the predicted jet drop distribution for the cases (c, d) may be missing some drops below radius $r_d$ = 1~\textmu{m}, which can be seen from the gap between the predicted and measured distributions where the film and jet curves overlap. \citet{neel2021} and \citet{neel_surf_2022} observed a similar shift/broadening in the peak size of ejected drops for collective bursting of millimetric bubbles in surfactant solutions of various concentrations, which remains unexplained. These inconsistencies may be explained by a surfactant or collective effect and require further investigation. We note also that the predicted jet drop production is sensitive to the chosen critical Laplace number ($La^*$, the drop ejection threshold), shifting the predicted drop size distribution to somewhat smaller sizes if this value is increased. 

% summary of discussion
Overall, the drop size distribution produced by clustered bursting bubbles in artificial seawater can be reasonably well-described from knowledge of the bubble size distribution, especially for the broad-banded distributions. By integrating individual bubble bursting scaling laws for size, number, and distribution of ejected drops for all measured bubble sizes, we arrive at a reasonable estimate of the drop size distribution. Further study is required to explain the details of deviations in large jet drop production and spray generation by denser rafts of small bubbles, and to more precisely connect the two production mechanisms around their respective cutoffs around 1~\textmu{m} (large size for film drops, and small size for jet drops), with possible collective effects. 
\section{Conclusion}
\label{sec:concl}

In this paper, we have characterized drop production by collective bubble bursting using a steady-state bubbling tank setup. Systematic measurements of bulk bubbles, surface bubbles, and drops/salt aerosols from many combined techniques allowed us to measure bubble radii spanning 30\textendash5000~\textmu{m} and drop radii spanning 0.05\textendash500~\textmu{m}, covering five total orders of magnitude that correspond to realistic ocean scales for drop production by bursting of bubbles generated by breaking waves. We used the complete measured size distributions to analyze the physical link between bubbles and drops for a variety of cases with different bulk bubble size distributions, both by directly attributing measured drops to measured bubbles and by applying scaling laws for individual bubble bursting to predict the drop size distributions from the measured bubble size distributions. 

By analyzing collective bubble bursting for a variety of bulk bubble size distributions in the experimental setup presented in this paper, we have observed that:

\begin{enumerate}[(i)]
    \item Bulk and surface bubbles can be reasonably linked through a simple flux idea assuming a bulk bubble rise velocity parameterization and constant surface bubble lifetime, and coalescence is not a dominant effect to consider with artificial seawater (figure \ref{fig:flux}).
    \item Bursting bubbles with radii greater than $\sim$1~mm do not produce significant numbers of submicron drops (shown most clearly in figure \ref{fig:invattr}(a-f)).
    \item A modest collective effect (around a factor 2) may decrease the production efficiency of clustered bubbles compared to individual bubble bursting (figure \ref{fig:invattr}(g-i)). 
    \item Scaling laws developed for individual bubble bursting can describe the measured drop size distribution well for broad-banded bulk bubble size distributions characterized by a continuum of both bubble and ejected drop sizes (figures \ref{fig:attr4lines}(a) and \ref{fig:fwd}(a, b)). Drops predicted by the proposed film flapping mechanism for bubbles of radii 70\textendash1000~\textmu{m} describe the submicron portion of the drop size distribution with an efficiency of about 40 drops per bubbles. Drops above approximately 1~\textmu{m} can be explained following scaling laws for jet drops with an efficiency for the number of drops produced by bubbles of radii 30\textendash2300~\textmu{m} between 3\textendash0.6 jet drops per bubble.
    \item Cases with a narrower size range of bulk bubbles were not described as well by the single bubble scalings. Jet drop scalings primarily overpredicted the production of large drops (and displayed a shift in size) compared to the experimental measurements. This discrepancy remains unexplained but suggests the possibility of a collective effect \textendash\ when clusters of bubbles of similar sizes burst \textendash\ that may modify bursting efficiency or the link between sizes of bubbles and ejected drops (figures \ref{fig:attr4lines}(b) and \ref{fig:fwd}(c, d)).
\end{enumerate}

Overall, we have studied the link between bulk bubbles, surface bubbles, and drops in a collective bursting setup, and we have experimentally confirmed the applicability of mechanistic ocean spray emissions functions based on individual bursting mechanisms \citep{deike_rev_2022,deike_mechanistic_2022}. We have leveraged many combined measurements and cases of diverse bubble size distributions and underwater conditions to analyze the attribution of drops to bursting bubbles. The potential collective effect suggested by these experiments will be studied further through dynamical measurements of bursting in bubble rafts. 
Future work using the same setup will explore how the link between bursting bubbles and emitted drops changes for solutions of different surfactant and temperatures, further extending the applicability of physics-based sea spray emissions functions \citep{deike_mechanistic_2022} to open ocean conditions in presence of biological activity \citep{burrows_oceanfilms_2014} and taking steps to understand sea spray emissions by bubble bursting in other increasingly realistic configurations for ocean conditions. 

\vspace{10pt} 

\noindent\textbf{Funding:}
This work was supported by the National Science Foundation under grant numbers 2122042, 1849762, 1844932, 2318816 to LD; the Cooperative Institute for Modeling the Earth's System at Princeton University; the National Science Foundation Graduate Research Fellowship to MM; and the High Meadows Environmental Institute Hack award to MM.

\vspace{5pt} 
%\textbf{Declaration of Interests}
\noindent The authors report no conflict of interest.

\vspace{5pt}
\noindent\textbf{Data availability statement:}
Data used to prepare this work is publicly available at 
\url{https://doi.org/10.34770/7s70-0v84}.

\vspace{5pt} 

\noindent\textbf{Author ORCIDs:}
M. Mazzatenta, https://orcid.org/0000-0002-8362-0349; M.A. Erinin, https://orcid.org/0000-0002-1660-6002; B. N\'eel, https://orcid.org/0000-0002-2218-5945; L. Deike, https://orcid.org/0000-0002-4644-9909.

\newpage
\appendix

\section{Average drop production efficiencies}\label{appA}

\begin{figure}[h!]
  \centerline{\includegraphics[scale=.9]{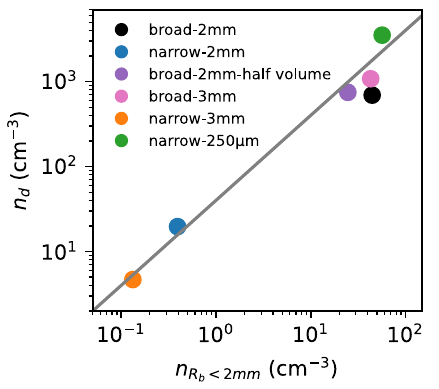}}
  \caption{Total number of drops ($n_d$) versus the number of bubbles with radii below 2~mm ($n_{R_b<2mm}$), both per unit volume, for each case explored in the paper. The gray line represents $n_d = 40~n_{R_b<2mm}$, where the coefficient represents an average drop production efficiency, assuming drop production by bubbles with $R_b <$ 2~mm.}
\label{fig:numbers}
\end{figure}

The different conditions explored throughout the paper are characterized by very different numbers of bubbles at the surface, and therefore very different numbers of emitted drops, as illustrated by figure \ref{fig:numbers}, showing the number of drops as a function of the number of bubbles with radii smaller than 2~mm. As indicated in the figure and already mentioned, our experiments span a range of number of bubbles around three orders of magnitude. An averaged drop production efficiency for each surface bubble configuration can be evaluated by an approximate linear relationship between the number of drops and the number of bubbles, as indicated by the solid line in figure \ref{fig:numbers}. This linear fit suggests similar average efficiencies among all cases, with about 40 drops per bursting bubble. We also observe that the two cases of broad-banded bubble size distributions with significant clustering (cases Broad-2~mm and Broad-3~mm) show a reduced efficiency (pink and black symbols being below the linear trend). 

Experiments were also performed for various air flowrates of 200~cm$^3$/min, 400~cm$^3$/min, and 800~cm$^3$/min for the bubbler configuration, and all cases yield drop production efficiencies similar to those reported in figure \ref{fig:numbers}, with no significant measurable effect of the air flowrate on the production of small drops in our setup.

\section{Uncertainty in drop production efficiencies and sensitivity of predicted drop size distributions}\label{appB}

When performing the direct attribution of drops to bubbles from the data (section \ref{sec:direct}) and calculating drop production efficiency for film and jet drops, there are uncertainties in the measurement of the bubble size distribution and drop size distribution as well as uncertainties related to the chosen radius bounds of integration. As discussed in the experimental methods, the experimental size distributions are statistically converged, and the statistics are highly reproducible when repeating the same experiment. From measurements of bubble and drop size distributions, the estimated uncertainty on the efficiency values shown in figure \ref{fig:invattr} is around a factor 2. 

We note, however, that the predicted drop size distributions shown in sections \ref{section:forward} and \ref{sec:fwd2} (or the figure \ref{fig:invattr} efficiencies calculated in a particular size bucket) can be quite sensitive to the choice of bubble and/or drop radius ranges that are integrated for each drop production mechanism (film or jet). To calculate the efficiencies presented in figure \ref{fig:invattr}, the total ranges of assumed jet drops and corresponding bubbles are split into approximately equal log-spaced buckets, of proportional sizes between the drop and bubbles. Shrinking/expanding/shifting these buckets will cause the efficiency values to shift up or down slightly, which could change the physical interpretation. The interpretation we propose results from our understanding of individual bursting processes discussed in the literature and from the consistency between the message of the direct attribution (section \ref{sec:direct}) and the drop size distributions predicted by the individual bubble bursting scaling laws combined with the measured bubble size distributions (sections \ref{section:forward}, \ref{sec:fwd2}). We note that the overall average drop production efficiencies (see appendix \ref{appA}) are relatively constant across all cases. This overall efficiency is dominated by small drop production by bubbles of intermediate size, which is shown in both our methods of drop attribution to bursting bubbles for all cases, and gives confidence in our chosen interpretation.

\begin{figure}[h!]
  \centerline{\includegraphics[scale=.6]{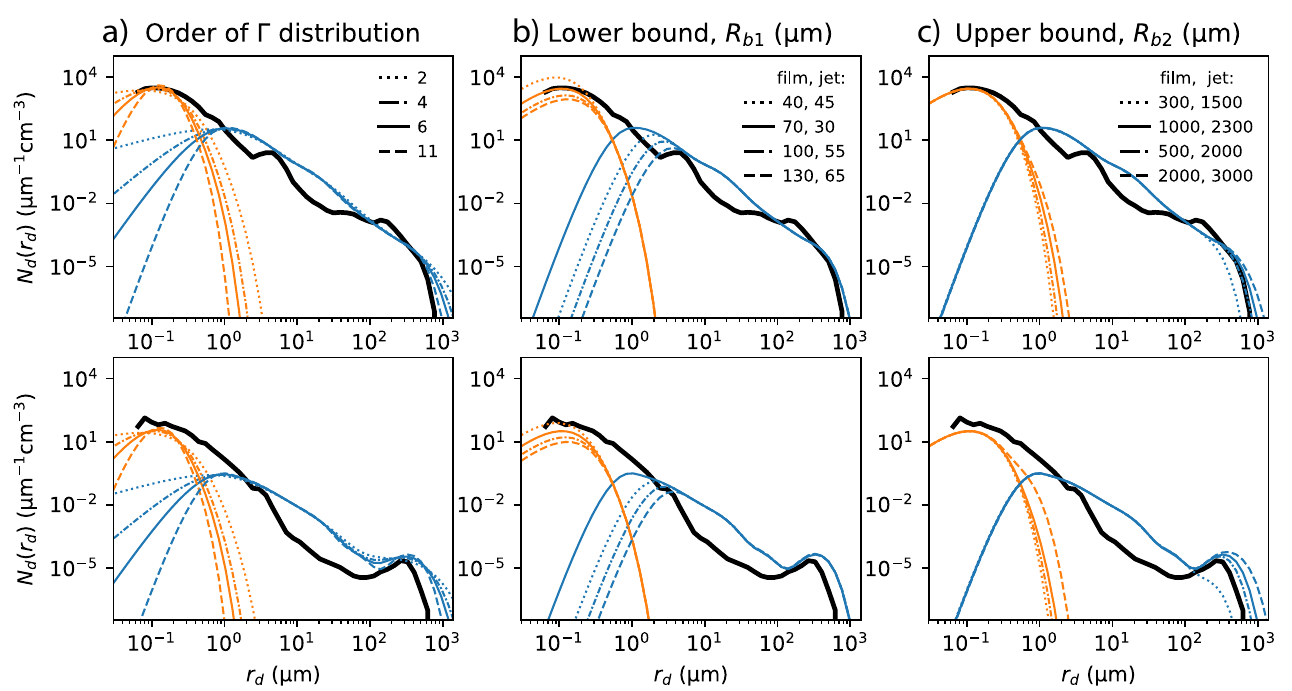}}
  \caption{Bursting bubble ($N_b(R_b)$) and drop ($N_d(r_d)$) size distributions for the Broad-2~mm (top row) and Narrow-2~mm (bottom row) cases. The predicted drop size distributions from individual bubble bursting scaling laws are plotted in color for film flapping (orange) and jet (blue, from G-C scaling) drop production. The sensitivity of the predicted drop size distributions to parameters of the model/integration (equation \ref{globaldrop}) is shown for: (a) the order of the Gamma distribution $p(r_d/\langle r_d \rangle, R_b)$ (varied between 2 and 11), (b) the lower bound, $R_{b1}$, for the selected radius range (varied between 40 and 130~\textmu{m} for film flapping and between 30 and 65~\textmu{m} for jet), and (c) the upper bound, $R_{b2}$, for the selected radius range (varied between 300 and 2000~\textmu{m} for film flapping and between 1500 and 3000~\textmu{m} for jet). For the lower and upper bounds, the solid line represents the distributions/set of parameters used in figure \ref{fig:attr4lines}. Parameters are described further in table \ref{tab:paramsfull}).}
\label{fig:sensitivity}
\end{figure}

Regarding the analysis presented in sections \ref{section:forward} and \ref{sec:fwd2}, we provide further details on the sensitivity of the drop size distributions predicted using the single bubble scalings (used to generate the colored curves presented in figures \ref{fig:attr4lines} and \ref{fig:fwd}), shown in figure \ref{fig:sensitivity}. We test the sensitivity to (a) the order of the Gamma distribution representing $p(r_d/\langle r_d \rangle, R_b)$ and to (b) the lower and (c) upper radius bounds chosen for the film flapping and jet drop production mechanisms. From these figures, we observe that the order of the Gamma distribution primarily affects the shape of the curve at the small drop-size cutoff, and the chosen upper radius bound shifts the large drop-size cutoff. The predicted distributions are much more sensitive to the chosen lower radius bound for each production mechanism, which more significantly affects the magnitude of the predicted distributions for the film flap drops and is critical in defining the location of the peak for small jet drops (and therefore also the important overlap between film flap and jet scalings). 

We note that we use the Gamma distribution to represent the statistics of drops emitted by a single bubble bursting, but another type of distribution, such as a lognormal distribution, could be used. As discussed in \citet{berny2021}, when integrating over a wide range of bubbles sizes, the scaling for the mean size and number of drops emitted dominates over the type of function used, and the type of function will mostly modulate the result near the bounds of integration. The Gamma distribution has the practical advantage of allowing for analytical integration when considering bubble size distributions represented as power laws, as discussed in \citet{lv2012}.

\begin{figure}[h!]
  \centerline{\includegraphics[scale=.6]{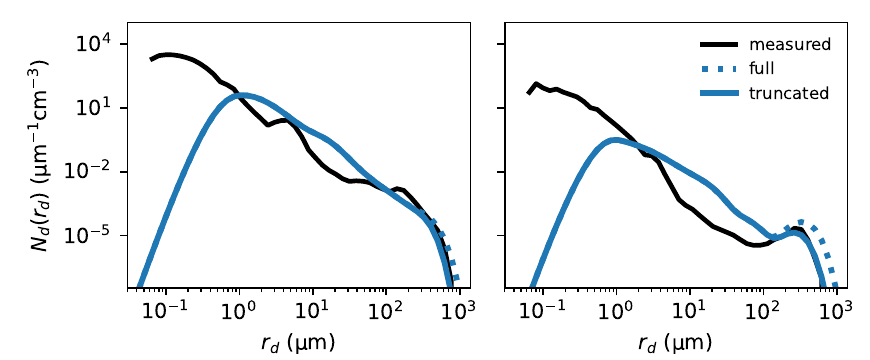}}
  \caption{Measured and predicted drop size distributions ($N_d(r_d)$) for the Broad-2~mm (left) and Narrow-2~mm (right) cases. The predicted jet drop size distributions are computed using the full Gamma distribution for all bubble radii  (dotted line, also shown in figure \ref{fig:attr4lines}) and the truncated Gamma distribution (solid line). The truncated Gamma distribution is equal to zero for bubbles of $R_b >$ 2~mm.}
\label{fig:truncated}
\end{figure}

Regarding the larger drop sizes, we can also consider using a truncated Gamma distribution in the integration. In this case, the distribution of drop sizes produced by an individual bursting bubble ($p(r_d/\langle r_d \rangle, R_b)$) would be represented by the Gamma distribution for bubbles smaller than a chosen cutoff radius $R_b^{max}$, and equal to zero for bubbles larger than this cutoff size. Figure \ref{fig:truncated} shows the comparison between jet drop size distributions computed using the full Gamma distribution and the truncated version, which is set to zero for bubbles larger than the chosen cutoff radius of $R_b^{max}$ = 2~mm. Using this truncated Gamma distribution, the predicted drop size distributions describe the data for the large drop sizes very well. 

\newpage

\bibliographystyle{biblio}
\bibliography{biblio}

\end{document}